\documentclass[final,twoside,11pt]{entics} 
\usepackage{enticsmacro}
\usepackage{graphicx}
\usepackage[all]{xy}
\usepackage{appendix}
\usepackage[pass]{geometry}
\usepackage{amsfonts}
\usepackage{hyperref}
\usepackage{amsmath,xspace}
\usepackage{amssymb}
\usepackage{bm}
\usepackage{mathpartir}
\usepackage{mathtools}
\usepackage{thm-restate}
\usepackage{relsize}
\usepackage{bussproofs}
\usepackage{stmaryrd}
\usepackage{tikz}
\usetikzlibrary{matrix}
\usetikzlibrary{decorations.pathreplacing,calligraphy}
\usetikzlibrary{decorations.pathmorphing}
\usepackage[framemethod=tikz]{mdframed}
\mdfsetup{
    innertopmargin=-5pt,
    innerleftmargin=4pt,
    innerrightmargin=4pt
}
\usepackage{centernot}
\usepackage{xcolor}
\usepackage{cleveref}
\usepackage{crossreftools}

\usepackage{thmtools}
\usepackage{etoolbox}

\definecolor{cblRed}{rgb}{.60,.00,.01}
\definecolor{cblBlue}{rgb}{.16,.23,.42}
\definecolor{cblBlueLt}{rgb}{.00,.56,.61}
\definecolor{cblYellow}{rgb}{1.00,.69,.00}
\definecolor{cblPink}{rgb}{.85,.00,.42}
\definecolor{cblPurple}{rgb}{.10,.00,.54}
\definecolor{cblPurpleLt}{rgb}{.76,.39,.89}
\definecolor{cblOrange}{rgb}{.74,.29,.00}
\definecolor{cblGreen}{rgb}{.38,.61,.27}

\newcommand{\clrone}[1]{\textcolor{cblRed}{#1}}
\newcommand{\clrtwo}[1]{\textcolor{cblGreen}{#1}}
\newcommand{\pctxcolor}{cblBlueLt}

\newcommand{\nconf}{non-con\-flu\-ent\xspace}

\newcommand{\myconf}{con\-flu\-ent\xspace}

\newcommand{\col}{\nconf}

\newcommand{\nond}{non-de\-ter\-min\-ism\xspace}
\newcommand{\nondt}{non-de\-ter\-min\-is\-tic\xspace}

\newcommand{\lamcoldetsh}{\ensuremath{\lambda_{\mathtt{C}}}\xspace}

\newcommand{\core}[1]{#1^{\downarrow}}

\newcommand{\linvar}[1]{ #1 ^{\ell}}
\newcommand{\unvar}[1]{ #1 ^{!}  }
\newcommand{\lunvar}[1]{#1}
\newcommand{\bagsep}{\star}

\newcommand{\sep}{\ | \ } 
\newcommand{\bag}[1]{\lbag #1 \rbag}
\renewcommand{\1}{\mathtt{1}}
\newcommand{\oneb}{\mathtt{1}}

\newcommand{\dom}[1]{\mathit{dom}(#1)}
\newcommand{\arrt}[2]{\ensuremath{#1 \rightarrow #2}}
\newcommand{\fail}{\mathtt{fail}}

\newcommand{\concat}{\ensuremath{\diamond}}
\newcommand{\relunbag}{\ensuremath{\sim}}
\newcommand{\sharing}[2]{[#1 \leftarrow #2]}

\newcommand{\size}[1]{\mathsf{size}(#1)} 
\newcommand{\headf}[1]{\ensuremath{\mathsf{head}(#1)}}

\newcommand{\dash}{\text{-}}

\def\esubst#1#2{\langle\!\langle \raisebox{.5ex}{\small$#1$}\! / \mbox{\small$#2$}\rangle\!\rangle} 

\newcommand{\headlin}[1]{  \{ #1 \} } 
\newcommand{\ltalltriangle}{{\langle\!|}}
\newcommand{\rtalltriangle}{{|\!\rangle}}
\newcommand{\linexsub}[1]{\ltalltriangle #1 \rtalltriangle}
\newcommand{\unexsub}[1]{\llfloor #1 \rrceil }

\newcommand{\lfv}[1]{\mathsf{fv}(#1)}
\newcommand{\llfv}[1]{\mathsf{lfv}(#1)}

\newcommand{\unit}{\mathbf{unit}}
\newcommand{\head}[1]{\mathsf{head}(#1)} 

\newcommand{\red}{\mathbin{\longrightarrow}}
\newcommand{\redone}{\ensuremath{\mathbin{\clrone{\longrightarrow}}}\xspace}
\newcommand{\redtwo}{\ensuremath{\mathbin{\clrtwo{\leadsto}}}\xspace}

\newcommand{\readyPrefixBisimSub}[1]{\mkern-1mu\mathtt{#1}}
\newcommand{\readyPrefixBisim}[1]{\sim_{\readyPrefixBisimSub{#1}}}
\newcommand{\nreadyPrefixBisim}[1]{\not\sim_{\readyPrefixBisimSub{#1}}}

\newcommand{\succone}{\mathbin{\clrone{\Downarrow}}}

\newcommand{\redlab}[1]{\ensuremath{\mathtt{[#1]} }}

\newcommand{\defref}[1]{Def.\ \labelcref{#1}\xspace}

\newcommand{\wfdash}{\vDash}
\newcommand{\wtdash}{\vdash}

\newcommand{\some}{\mathtt{some}}
\newcommand{\none}{\mathtt{none}}

\newcommand{\fn}[1]{\mathit{fn}(#1)}
\newcommand{\fln}[1]{\mathit{fln}(#1)}

\newcommand{\ampy}{\mathbin{\bindnasrepma}}
\newcommand{\with}{{\binampersand}}
\newcommand{\onef}{\mathbf{1}}
\newcommand{\dual}[1]{\overline{#1}}

\newcommand{\encod}[2]{\llbracket#1\rrbracket_{#2}} 

\newcommand{\piencodf}[1]{{\textcolor{cblPink}{\llbracket}#1\textcolor{cblPink}{\rrbracket}}}
\newcommand{\piencod}[1]{\piencodf{#1}}
\newcommand{\piencodfscale}[1]{\color{cblPink}\left\llbracket \normalcolor #1 \color{cblPink}\right\rrbracket \normalcolor}

\newcommand{\succp}[2]{\ensuremath{{#1} \Downarrow {#2}}}

\newcommand{\premat}{\succeq_{\mkern-2mu \scalebox{.8}{$\nd$}} }
\newcommand{\premattwo}{\preceq_{\nd} }

\newcommand{\relalpha}{ \bowtie }

\newcommand{\readyPrefix}[1]{\downarrow_{#1}}

\newcommand{\bignd}{\scalebox{1.5}{$\nd\mkern-3mu$}}

\newcommand{\ol}[1]{\overline{#1}}
\newcommand{\0}{\bm{0}}
\newcommand{\fwd}{\mathbin{\leftrightarrow}}
\newcommand{\sepr}{\mathbin{\scalebox{1.3}{$\mid$}}}

\newcommand{\rredone}[1]{[\clrone{\shortrightarrow}_{#1}]}
\newcommand{\rredtwo}[1]{[\clrtwo{\rightsquigarrow}_{#1}]}
\newcommand{\ttype}[2][]{\text{\normalfont#1[\scc{T#2}]}}

\newcommand{\sff}[1]{\relax\ifmmode\mathsf{#1}\else\textsf{#1}\fi}
\newcommand{\mathsc}[1]{\text{\normalfont\scshape#1}}
\newcommand{\scc}[1]{\relax\ifmmode\mathsc{#1}\else\textsc{#1}\fi}

\newcommand{\tensor}{\ensuremath{\mathbin{\otimes}}}
\newcommand{\parr}{\mathbin{\rott{\with}}}
\newcommand{\rott}[1]{\mathpalette\rot{#1}}
\newcommand{\rot}[2]{\rotatebox[origin=c]{180}{$#1{#2}$}}

\newcommand{\D}[1]{\textcolor{\pctxcolor}{\llparenthesis}\mkern1mu #1 \mkern1mu\textcolor{\pctxcolor}{\rrparenthesis}}

\newcommand{\mhole}{[\cdot]}

\let\obigcup\bigcup
\renewcommand{\bigcup}{{\mathchoice{\textstyle}{}{}{}\obigcup}}
\let\osum\sum
\renewcommand{\sum}{{\mathchoice{\textstyle}{}{}{}\osum}}
\let\oprod\prod
\renewcommand{\prod}{\mathchoice{\textstyle}{}{}{}{\oprod}}

\newcommand{\prlplus}{\hbox{\rlap{$\mkern-.4mu{}\mathbin{|\mkern-3mu|}{}$}$-$}}
\newcommand{\nd}{\mathbin{\prlplus}}

\newcommand{\piprecong}[1]{\mathrel{\succeq_{#1}}}
\newcommand{\sub}[1]{\ensuremath{\mathit{subj}\{#1\}}}

\newcommand{\mtt}[1]{\mathtt{#1}}
\let\oprl\|
\renewcommand{\|}{\mathbin{|}}
\newcommand{\pclose}[1]{\ol{#1}[]}
\newcommand{\gclose}[1]{#1()}
\newcommand{\psome}[1]{\ol{#1}.\some}
\newcommand{\pnone}[1]{\ol{#1}.\none}
\newcommand{\gsome}[2]{{#1.\some}_{#2}}
\newcommand{\pname}[2]{\ol{#1}[#2]}
\newcommand{\puname}[2]{{?}\ol{#1}[#2]}
\newcommand{\gname}[2]{#1(#2)}
\newcommand{\guname}[2]{{!}#1(#2)}
\newcommand{\psel}[2]{\ol{#1}.#2}
\newcommand{\gsel}[1]{#1.\mtt{case}}
\newcommand{\pfwd}[2]{[#1 \fwd #2]}
\newcommand{\res}[1]{(\bm{\nu} #1)}

\newcommand{\clpi}{\ensuremath{{{\mathsf{s}\pi\mkern-2mu}^{+}\mkern-1mu}}\xspace}
\newcommand{\fullclpi}{\ensuremath{{{\mathsf{s}\pi\mkern-1mu}^{!}\mkern-1mu}}\xspace}

\newcommand{\bn}[1]{\mathit{bn}(#1)}
\newcommand{\fpn}[1]{\mathit{fpn}(#1)}

\newcommand{\dashes}{\vspace{-1.25em}\hbox to \textwidth{\leaders\hbox to 3pt{\hss . \hss}\hfil}\smallskip}
\newcommand{\mcl}[1]{\mathcal{#1}}

\newcommand{\lctx}[1]{\mcl{#1}}
\NewDocumentCommand \pctx {s O{} m o} {\IfBooleanT{#1}{\textcolor{\pctxcolor}{\llparenthesis}\mkern1mu}\textcolor{\pctxcolor}{\mathtt{#3}}\IfBooleanT{#1}{\mkern1mu\textcolor{\pctxcolor}{\rrparenthesis}} \IfValueT{#4}{\textcolor{\pctxcolor}{#2[}#4\textcolor{\pctxcolor}{#2]}}}
\newcommand{\sucs}[1]{{\checkmark\mkern-4mu}_{#1}}

\def\shred{\mathbin{\shortrightarrow}}
\def\ndots{\hbox to 1em{.\hss.\hss.}}

\newcommand{\myDots}{\ifmmode\mathinner{\ldotp\kern-0.2em\ldotp\kern-0.2em\ldotp}\else.\kern-0.13em.\kern-0.13em.\fi}

\newcommand{\myrev}[1]{#1}
\newcommand{\myrevopt}[1]{}
\newcommand{\mysmall}{\small}

\def\conf{MFPS 2024}
\def\myconf{mfps40} 	
\volume{4}			
\def\volu{4}			
\def\papno{12}			
\def\mydoi{14735}%
\def\lastname{Van den Heuvel et al} 
\def\runtitle{Typed Non-determinism in Concurrent Calculi} 
\def\copynames{B. van den Heuvel, D. Nantes-Sobrinho,}
\def\copyname{\  \ J.W.N. Paulus, J.A. P\'erez\ }    
\begin{document}
\begin{frontmatter}
  \title{Typed Non-determinism in Concurrent Calculi: \\[1ex] The Eager Way}
\author{Bas van den Heuvel\thanksref{a}
}
  \author{Daniele Nantes-Sobrinho\thanksref{b}
  }
  \author{Joseph W.N. Paulus \thanksref{c}
  }	
   \author{Jorge A. P\'erez\thanksref{d}
   }		
     \address[a]{Karlsruhe University of Applied Sciences, Karlsruhe, and University of Freiburg, Freiburg, Germany} 
   \address[b]{Department of Computing, Imperial College London,				
    London, UK}  							
  \address[c]{University of Oxford, Oxford, UK} 
 \address[d]{University of Groningen, Groningen,  The Netherlands} 
\begin{abstract} 
We consider the problem of designing typed concurrent calculi with \emph{non-deterministic choice} in which types leverage \emph{linearity} for controlling resources, thereby ensuring strong correctness properties for processes. 
This problem is constrained by the delicate tension between non-determinism and linearity. 
Prior work developed a session-typed $\pi$-calculus with standard non-deterministic choice; well-typed processes enjoy type preservation and deadlock-freedom.
Central to this typed calculus is a \emph{lazy} semantics that gradually discards branches in  choices.
This lazy semantics, however,   is complex: various technical elements are needed to  describe the non-deterministic behavior of typed processes. 
This paper develops an entirely new approach, based on an \emph{eager} semantics, which more directly represents choices and commitment. 
We present a $\pi$-calculus in which non-deterministic choices are governed by 
this eager semantics  and session types. 
We establish its key correctness properties, including deadlock-freedom, 
and demonstrate its expressivity by  correctly translating a typed 
resource $\lambda$-calculus.
\end{abstract}
\begin{keyword}
Concurrency, process calculi, linear type systems, session types, intersection types, non-determinism.
\end{keyword}
\end{frontmatter}

\section{Introduction}\label{intro}

This paper addresses the problem of designing concurrent calculi with \emph{non-deterministic choice} in which types leverage \emph{linearity} for controlling resources. 
Specifically, our interest is in variants of the $\pi$-calculus, the paradigmatic calculus of concurrency and interaction~\cite{DBLP:journals/iandc/MilnerPW92a,DBLP:books/daglib/0004377}; here  the resources are the names (or channels) that communicating processes use to perform protocols described by \emph{session types}~\cite{DBLP:conf/concur/Honda93,DBLP:conf/esop/HondaVK98}. 
This is a challenging design problem, due to the delicate tension between non-determinism and linearity. 
On the one hand, resource control based on linearity is essential to statically enforce important correctness properties for processes: \emph{protocol fidelity} (processes respect their protocols), \emph{communication safety} (processes never incur into message mismatches), and \emph{deadlock-freedom} (processes never get stuck). 
On the other hand, implementing the usual (non-confluent) semantics of non-deterministic choice is at odds with linearity: a careless handling of discarded branches in choices can jeopardize resources meant to be used exactly once. 

To better understand the problem, 
it is instructive to recall the reduction rule for the (untyped) $\pi$-calculus (e.g.,~\cite{DBLP:books/daglib/0004377}):
\begin{equation}
  (\pname{x}{z};P_1 + M_1) \|
(\gname{x}{y};P_2 + M_2)
\longrightarrow
P_1 \| P_2\{ z / y \}
\label{eq:usualpi}
\end{equation}
Rule~\eqref{eq:usualpi} specifies the interaction between two (binary) choices: it coalesces a synchronization along name $x$ (whereby name $z$ is passed around) with the commitment to retaining the branches involved in the exchange. Indeed, after  reduction the two branches not involved in the synchronization, $M_1$ and $M_2$, are discarded.
While  appropriate in the untyped setting, it would be unwise to adopt this rule in a linearly-typed setting: clearly,  $M_1$ and $M_2$ could as well denote resources that must be used exactly once.

The core technical problem is then how to devise formulations for non-determinism that preserve the non-confluent character captured by Rule~\eqref{eq:usualpi}---which effectively expresses commitment in specifications---while respecting the principles of linearity-based resource control. 

As an answer to this problem, our prior work~\cite{DBLP:conf/aplas/HeuvelPNP23} introduced a typed process model with a new non-deterministic choice operator, denoted $P \nd Q$, in which  $P$ and $Q$ act upon the same  (linear) resources---they are  branches that denote  different implementations of the same session protocols.
This  choice operator is governed by a \emph{lazy} semantics that minimizes commitment as much as possible. 
Roughly speaking, the lazy semantics distinguishes between ``possible'' and ``impossible'' branches, depending on the synchronizations enabled in a process and its context. This distinction allows us to reduce the set of branches under consideration: the actual  choice   (as in Rule~\eqref{eq:usualpi}) is enacted at the level of possible branches. 

The lazy semantics meets our desiderata: it expresses commitment and respects linearity of resources. 
Also, the resulting typed process model   enforces the three correctness properties given above, and is also expressive enough to precisely encode a resource $\lambda$-calculus with non-deterministic behavior and explicit failures.
Still, the lazy approach is not entirely satisfactory: its definition is complex and so the behavior of non-deterministic choices cannot be easily discerned. 
The lazy semantics rests upon a pre-order on processes (which captures the intermediate distinction between possible and impossible branches) and a compatibility relation on prefixes; also, it needs to be indexed by the names involved in the synchronization. This required machinery is not ideal, in particular if one contrasts it with the compact and effective  Rule~\eqref{eq:usualpi}.

In this paper, we  devise a formulation of non-determinism that is simpler and more direct than the lazy semantics of~\cite{DBLP:conf/aplas/HeuvelPNP23}.
We propose an \emph{eager} semantics that enforces commitment by examining the contexts under which reductions occur. 
This is an economical solution, as it rests upon a simple definition that is arguably easier to understand and reason about than the lazy semantics. Perhaps more importantly, our new eager semantics still meets our desiderata on commitment and linearity as enforced by typing.

Clearly, giving an alternative eager semantics for the typed process model in~\cite{DBLP:conf/aplas/HeuvelPNP23} immediately raises the question of its positioning with respect to preceding developments. Several interesting issues arise. Does the eager semantics fit well with the session type system in~\cite{DBLP:conf/aplas/HeuvelPNP23}? 
Because the typed process model with the lazy semantics was shown to precisely encode a resource $\lambda$-calculus with non-determinism, we may also ask:  does moving to a simpler operational setting affect expressivity?  Moreover, how does the eager semantics compare to the lazy semantics, independently from the ability of encoding advanced $\lambda$-calculi?

This paper's goal is to provide technical answers to these questions. 
The base language for our new eager semantics is \fullclpi , the extension of the session $\pi$-calculus  in \cite{DBLP:conf/aplas/HeuvelPNP23} with unrestricted behaviors (client and server constructs). This way, our developments are based on a richer setting than in~\cite{DBLP:conf/aplas/HeuvelPNP23} (where only linear behaviors were considered). Having defined the eager semantics, we move to consider the associated session type system. We actually consider the exact same type system as in~\cite{DBLP:conf/aplas/HeuvelPNP23} and establish that well-typed processes satisfy the same properties (type preservation and deadlock-freedom). These results are reassuring: they confirm that our eager semantics does not break properties derived from typing, and that the eager/lazy distinction remains an  operational concern, which does not transpire at the level of typing.

We then assess the expressiveness of the eager process model by giving a process interpretation of \lamcoldetsh, a resource $\lambda$-calculus with non-determinism. Also in this case, we consider an extension of the language considered in~\cite{DBLP:conf/aplas/HeuvelPNP23}: the calculus \lamcoldetsh features both linear and unrestricted resources, which requires several innovations, in particular for the associated intersection type system (the $\lambda$-calculus in  \cite{DBLP:conf/aplas/HeuvelPNP23} is the sub-language of \lamcoldetsh  with  linear resources only). The translation of \lamcoldetsh  into \fullclpi we present here also features innovations: while its linear portion (i.e., the translation of terms with linear resources into linear processes) is the same as in \cite{DBLP:conf/aplas/HeuvelPNP23}, the translation of terms with unrestricted resources into client/server processes is new to this presentation. Again, this corroborates that the eager/lazy distinction is not relevant at the static level given by the translation. The salient differences appear at the \emph{dynamic} level, i.e., in the operational correspondence properties that relate the computations of a term in  \lamcoldetsh with the  behavior of its corresponding process in \fullclpi (and vice versa). Indeed, it turns out that the eager semantics of \fullclpi induces operational correspondences that are   ``looser'' than in the lazy regime. That is, the lazy semantics provides a tighter account of the dynamics of terms and their corresponding translations. 

In summary, our paper extends and complements the results in~\cite{DBLP:conf/aplas/HeuvelPNP23} with the following  contributions:

\begin{enumerate}
	\item \label{c:one} A new eager semantics for \fullclpi, the session $\pi$-calculus with  \nondt choice that extends the calculus introduced in \cite{DBLP:conf/aplas/HeuvelPNP23} with client/server behaviors (\Cref{ss:pisemantics}).
	\item A  type system for \fullclpi, which ensures type preservation and deadlock-freedom for well-typed processes governed by the eager semantics (\Cref{ss:piTypeSys}).
	\item The resource calculus \lamcoldetsh, which extends the one presented in \cite{DBLP:conf/aplas/HeuvelPNP23}  with unrestricted resources.
	Governed by intersection types, we establish subject reduction and subject expansion results (\Cref{s:lambda}).
	\item A typed translation of \lamcoldetsh into \fullclpi, with an analysis of its static and dynamic correctness (\Cref{s:transUnres}),
	and a comparison between our new eager semantics and the lazy semantics presented (\Cref{s:compare}).
\end{enumerate} 
\noindent
Omitted material can be found in~\cite{DBLP:journals/corr/abs-2205-00680}, which contains technical  details for both lazy and eager semantics. The PhD thesis of Paulus \cite{PaulusPhDThesis} also contains that omitted material and gives a complete treatment of other related translations of  typed $\lambda$-calculi into typed $\pi$-calculi. 
Throughout the paper we use different colors (such as \clrone{red}  and \clrtwo{green}) to improve readability. However, the paper can be followed in black-and-white.

\section{A Typed \texorpdfstring{$\pi$}{Pi}-calculus with Non-deterministic Choice}
\label{s:pi}

In this section, we start by giving the syntax of  \fullclpi, a session-typed $\pi$-calculus with  \nondt choice.
Following the linear calculus \clpi given in \cite{DBLP:conf/aplas/HeuvelPNP23}, the key feature in \fullclpi is the \nondt choice operator `$\!P \nd Q$'.
The key novelty is the eagerly committing semantics for `$\!\nd$', which is compatible with linearity (\Cref{ss:pisemantics}).
Following its predecessors~\cite{CairesP17,DBLP:conf/aplas/HeuvelPNP23}, 
we give a session type system for \fullclpi; intuitively, session types express protocols to be executed along channels.
We prove that well-typed processes under the new eager semantics satisfy two key properties: \emph{type preservation} and  \emph{deadlock-freedom}.

\begin{figure}[t]
    \begin{mdframed} \mysmall
        \begin{align*}
            P,Q &::=
            \0 & \text{inaction}
            & \quad \sepr
            \pfwd{x}{y} & \text{forwarder}
           &~~~\sepr
            P \| Q & \text{parallel}
            \\
            &~~~\sepr
            \res{x}(P \| Q) & \text{connect}
            & \quad \sepr
            P \nd Q & \text{non-determinism}
             & \quad \sepr
            \psome{x};P & \text{available}
            \\ &~~~\sepr
            \pname{x}{y};(P \| Q) & \text{output}
            & \quad \sepr
            \gname{x}{y};P & \text{input}
             & \quad \sepr
            \pnone{x} & \text{unavailable}
            \\ &~~~\sepr
            \psel{x}{\ell};P & \text{select}
            & \quad \sepr
            \gsel{x}\{i:P\}_{i \in I} & \text{branch}
            &~~~\sepr
            \gsome{x}{w_1,\ldots,w_n};P & \text{expect}
            \\ &~~~\sepr
            \puname{x}{y};P & \text{client request}
            & \quad \sepr
            \guname{x}{y};P & \text{server}
            \\ &~~~\sepr
            \pclose{x} & \text{close}
            & \quad \sepr
            \gclose{x};P & \text{wait}
        \end{align*}

        \smallskip
        \dashes

        \vspace{-5ex}
        \begin{align*}
            P &\equiv P' ~ [P \equiv_\alpha P']
            &
            \pfwd{x}{y} &\equiv \pfwd{y}{x}
            &
            P \| \0 &\equiv P
            \\
            (P \| Q) \| R &\equiv P \| (Q \| R)
            &
            P \| Q &\equiv Q \| P
            &
            \res{x}(P \| Q) &\equiv \res{x}(Q \| P)
            \\
            P \nd P &\equiv P
            &
            P \nd Q &\equiv Q \nd P
            &
            (P \nd Q) \nd R &\equiv P \nd (Q \nd R)
        \end{align*}
        \vspace{-6ex}
        \begin{align*}
            \res{x}((P \| Q) \| R) &\equiv \res{x}(P \| R) \| Q
            &
            [ x \notin \fn{Q} ]
            \\
            \res{x}(\res{y}(P \| Q) \| R) &\equiv \res{y}(\res{x}(P \| R) \| Q)
            &
            [ x \notin \fn{Q}, y \notin \fn{R} ]
            \\
            \res{x}(\guname{x}{y};P \| Q) &\equiv Q
            &
            [ x \notin \fn{Q} ]
        \end{align*}
    \end{mdframed}
    \caption{ \fullclpi: syntax (top) and structural congruence (bottom).}\label{f:pilang}
\end{figure}

\subsection{Syntax}

We use $P, Q, \ldots$ to denote processes, and $x,y,z,\ldots$ to denote \emph{names} representing  channels. 
\Cref{f:pilang} (top) gives the syntax of processes.
$P\{y/z\}$ denotes  the capture-avoiding substitution of $y$ for $z$ in  $P$.
Process~$\0$ denotes inaction, and
$\pfwd{x}{y}$ is a forwarder: a bidirectional link between $x$ and $y$.
There are two forms of parallel composition:
while the process $P \| Q$ denotes communication-free concurrency,
process $\res{x}(P \| Q)$ uses restriction $\res{x}$ to express that $P$ and $Q$ communicate   on  $x$ and do not share any other names.

Process $P\nd Q$ denotes the non-deterministic choice between $P$ and $Q$: intuitively, if one choice can perform a synchronization, the other option may be discarded if it cannot.
Since $\nd$ is associative, we often omit parentheses. 
Also, we write $\bignd_{i \in I} P_i$ for the non-deterministic choice between each $P_i$ for $i \in I$.

Our output construct integrates parallel composition and restriction: process $\pname{x}{y};(P \| Q)$ sends a fresh name $y$ along $x$ and then continues as $P\| Q$.
Types will ensure that behaviors on $y$ and $x$ are implemented by $P$ and $Q$, respectively, \myrev{which do not share any names; this is a form of communication-free concurrency that is key to avoiding deadlocks}.
The  input process  $\gname{x}{y};P$ receives a name $z$ along $x$ and continues as $P\{z/y\}$.
Process $\gsel{x}\{i:P_i\}_{i \in I}$  denotes a branch with labeled choices indexed by the finite set $I$: it awaits a choice on $x$ with continuation $ P_j$ for each $j \in I$.
The process $\psel{x}{\ell};P$ selects on $x$ the choice labeled $\ell$ before continuing as~$P$.
Processes $\pclose{x}$ and $\gclose{x}; P$ are dual actions for closing the session on~$x$.

Our language has server and client processes, not considered in~\cite{DBLP:conf/aplas/HeuvelPNP23}.
The server process $\guname{x}{y};P$ accepts requests from clients, receiving a name $z$ along $x$ to spawn $P\{z/y\}$; the server process remains available for further  requests.
A client request $\puname{x}{y};P$ sends a fresh name $y$ along $x$ and continues as $P$. Both client and server prefixes bind $y$ in $P$.

The remaining constructs define non-deterministic sessions which
may provide a protocol or fail~\cite{CairesP17}.
    Process $\psome{x}; P$ confirms the availability of a session on $x$ and continues as $P$.
        Process $\pnone{x}$ signals the failure to provide the session on $x$.
    Process $\gsome{x}{w_1,\ldots,w_n}
    ; P$ specifies a dependency on a non-deterministic session on $x$ (names $w_1, \ldots , w_n$ implement sessions in~$P$). This
    process can either (i) synchronize with a `$\psome{x}$' and continue as $P$, or (ii) synchronize
    with a `$\pnone{x}$', discard $P$, and propagate the failure to $w_1, \ldots , w_n$.
To reduce eye strain, in writing $\gsome{x}{}$ we freely combine names and sets of names.
This way, e.g.,
we write $\gsome{x}{y,\fn{P},\fn{Q}}$ rather than $\gsome{x}{\{y\} \cup \fn{P} \cup \fn{Q}}$.

Name $y$ is bound in $\res{y}(P \| Q)$, $\pname{x}{y};(P \| Q)$, and $\gname{x}{y};P$.
We write $\fn{P}$ and $\bn{P}$ for the free and bound names of $P$, respectively. The sets $\fln{P}$ and $\fpn{P}$ contain the free linear and non-linear names of $P$, respectively.
        Note that $\fpn{P} = \fn{P} \setminus \fln{P}$.
We adopt Barendregt's convention.

As usual, we shall use a \emph{structural congruence} ($\equiv$), the least congruence relation on processes induced by the rules in \Cref{f:pilang} (bottom).
Like the syntax of processes, the definition of $\equiv$ is aligned with the type system (defined next), such that $\equiv$ preserves typing (subject congruence, cf.\ Theorem~\labelcref{t:typePresEager}).
Notice that non-deterministic choice does not distribute over parallel and restriction.
The position of a non-deterministic choice in a process determines how it may commit, so changing its position affects commitment.

\subsection{The Lazy Semantics, by Example}
Reduction defines the steps that a process performs on its own.
A key contribution of our paper is an eager semantics for \fullclpi, which we present in \Cref{ss:pisemantics}. 
Before going into details, we find it instructive to illustrate the key ideas of the lazy semantics from~\cite{DBLP:conf/aplas/HeuvelPNP23}. 

The lazy semantics relies on 
   a precongruence on processes, denoted $\piprecong{S}$, where $S$ is a set that contains the names involved in a reduction: $S = \{x\}$ for a  synchronization on $x$, and  $S = \{x,y\}$ when a forwarder process $   \pfwd{x}{y}$ reduces.
   Building upon $\piprecong{S}$, the lazy semantics is then denoted $\redtwo_S$.  
   We  omit the curly braces; this way, e.g., we write `$\redtwo_{x,y}$' instead of `$\redtwo_{\{x,y\}}$'. The following example illustrates $\piprecong{S}$ and $\redtwo_{S}$:
   
\begin{example}
    \label{x:lazy}
    Consider a server that offers watching a movie's trailer or buying a movie using card or cash.
    We define the process  $\sff{MovieServer}_s := \gname{s}{\textit{title}} ; \sff{Movies}_s$, where $s$ is a name and $\sff{Movies}_s$ is as follows:
    \begin{align*}
        \sff{Movies}_s &:= \gsel{s} \{ \sff{buy} : \sff{MoviesBuy}_s , \sff{peek} : \sff{MoviesPeek}_s \}
        \\
        \sff{MoviesBuy}_s &:= \gsel{s}\{ \sff{card} : \sff{MoviesBuyCard}_s , \sff{cash} : \sff{MoviesBuyCash}_s \} 
        \\
        \sff{MoviesBuyCard}_s &:= \gname{s}{\textit{info}} ; \pname{s}{\texttt{movie}} ; \pclose{s}
        \\
        \sff{MoviesBuyCash}_s &:= \pname{s}{\texttt{movie}} ; \pclose{s} 
        \\
        \sff{MoviesPeek}_s &:= \pname{s}{\texttt{trailer}} ; \pclose{s}
    \end{align*}
    Now consider a client, Eve, undecided between buying `Barbie' or watching its trailer.
    If she decides to buy the movie, she cannot decide between paying with card or cash.
    We model Eve as follows, again using $s$ to communicate with the movie server, and modeling her indecisiveness with non-deterministic choices:
    \begin{align*}
        \sff{MovieClient}_s &:= \pname{s}{\texttt{Barbie}} ; \sff{Eve}_s 
        \\
        \sff{Eve}_s &:= \psel{s}{\sff{buy}} ; \psel{s}{\sff{card}} ; \sff{EveBuyCard}_s \nd \psel{s}{\sff{buy}} ; \psel{s}{\sff{cash}} ; \sff{EveBuyCash}_s \nd \psel{s}{\sff{peek}} ; \sff{EvePeek}_s
        \\
        \sff{EveBuyCard}_s &:= \pname{s}{\texttt{visa}} ; \gname{s}{\textit{movie}} ; \gclose{s} ; \0
        \\
        \sff{EveBuyCash}_s &:= \gname{s}{\textit{movie}} ; \gclose{s} ; \0
        \\
        \sff{EvePeek}_s &:= \gname{s}{\textit{link}} ; \gclose{s} ; \0
    \end{align*}
    
    We compose our movie server and Eve on $s$.
    Initially, the movie title is sent without non-determinism; notice how the reduction is annotated with `$s$' to indicate that the exchange occurred on   $s$:
    \[
        \res{s} ( \sff{MovieServer}_s \| \sff{MovieClient}_s )
        \redtwo_s
        \res{s} ( \sff{Movies}_s \| \sff{Eve}_s )
    \]
    At this point, two communications on $s$ are available: the selection of \sff{buy} and the selection of \sff{peek}.
    We model Eve's choice using the precongruence $\piprecong{s}$:
    \begin{mathpar}
        \sff{Eve}_s \piprecong{s} \psel{s}{\sff{buy}} ; \psel{s}{\sff{card}} ; \sff{EveBuyCard}_s \nd \psel{s}{\sff{buy}} ; \psel{s}{\sff{cash}} ; \sff{EveBuyCash}_s 
        \and
        \text{and}
        \and
        \sff{Eve}_s \piprecong{s} \psel{s}{\sff{peek}} ; \sff{EvePeek}_s
        .
    \end{mathpar}
    Notice how in the \sff{buy}-case the precongruence preserves the choice between the two methods of payment, because both choices start with the same selection.
    As such, performing the \sff{buy}-selection preserves this choice, whereas performing the \sff{peek}-selection results in a single alternative:
    \[
        \res{s} ( \sff{Movies}_s \| \sff{Eve}_s ) \redtwo_s \res{s} \big( \sff{MoviesBuy}_s \| ( \psel{s}{\sff{card}} ; \sff{EveBuyCard}_s \nd \psel{s}{\sff{cash}} ; \sff{EveBuyCash}_s ) \big)
    \]
    and
    \(
        \res{s} ( \sff{Movies}_s \| \sff{Eve}_s ) \redtwo_s \res{s} ( \sff{MoviesPeek}_s \| \sff{EvePeek}_s )
        .
    \)
    
    After the \sff{buy}-selection, the choice cannot be preserved, because the branches start with different selections; this is reflected by the precongruence on Eve's process:
    \[
        \psel{s}{\sff{card}} ; \sff{EveBuyCard}_s \nd \psel{s}{\sff{cash}} ; \sff{EveBuyCash}_s \piprecong{s} \psel{s}{\sff{card}} ; \sff{EveBuyCard}_s  
    \]
    and
    \(
        \psel{s}{\sff{card}} ; \sff{EveBuyCard}_s \nd \psel{s}{\sff{cash}} ; \sff{EveBuyCash}_s \piprecong{s} \psel{s}{\sff{cash}} ; \sff{EveBuyCash}_s
        .
    \)
    
    As such, after the \sff{buy}-selection, two further  reduction paths are possible. We give one of them:
    \[
        \res{s} \big( \sff{MoviesBuy}_s \| ( \psel{s}{\sff{card}} ; \sff{EveBuyCard}_s \nd \psel{s}{\sff{cash}} ; \sff{EveBuyCash}_s ) \big) \redtwo_s \res{s} ( \sff{MoviesBuyCard}_s \| \sff{EveBuyCard}_s )
    \]
\end{example}

\subsection{The New Eager Semantics}
\label{ss:pisemantics}

\begin{figure}[t]
    \begin{mdframed} \mysmall
        \begin{mathpar}
            \mathsmaller{\rredone{\scc{Id}}}~~
            \inferrule{}{
                \res{x}(\pctx[\big]{N}[\pfwd{x}{y}] \| Q) \redone \pctx*{N}[Q\{y/x\}]
            }
            \and
            \mathsmaller{\rredone{\bm{1} \bot}}~~
            \inferrule{}{
                \res{x}(\pctx{N}[\pclose{x}] \| \pctx{N'}[\gclose{x};Q]) \redone \pctx*{N}[\0] \| \pctx*{N'}[Q]
            }
            \and
            \mathsmaller{\rredone{\tensor \parr}}~~
            \inferrule{}{
                \res{x}(\pctx{N}[\pname{x}{y};(P \| Q)] \| \pctx{N'}[\gname{x}{z};R]) \redone \pctx*[\big]{N}[\res{x}(Q \| \res{y}(P \| \pctx*{N'}[R\{y/z\}]))]
            }
            \and
            \mathsmaller{\rredone{\oplus \with}}~~
            \inferrule{}{
                \forall k' \in K.~ \res{x}(\pctx{N}[\psel{x}{k'};P] \| \pctx{N'}[\gsel{x}\{k:Q^k\}_{k \in K}]) \redone \res{x}(\pctx*{N}[P] \| \pctx*{N'}[Q^{k'}])
            }
            \and
            \mathsmaller{\rredone{{?}{!}}}~~
            \inferrule{}{
                \res{x}(\pctx{N}[ \puname{x}{y};P ] \| \pctx{N'}[ \guname{x}{z};Q ])
                \redone \pctx*[\big]{N'}[ \res{x}\big( \res{y}( \pctx*{N}[P] \| Q\{y/z\} ) \| \guname{x}{z};Q \big) ]
            }
            \and
            \mathsmaller{\rredone{\some}}~~
            \inferrule{}{
                \res{x}(\pctx{N}[\psome{x};P] \| \pctx{N'}[\gsome{x}{w_1, \ldots, w_n};Q]) \redone \res{x}(\pctx*{N}[P] \| \pctx*{N'}[Q])
            }
            \and
            \mathsmaller{\rredone{\none}}~~
            \inferrule{}{
                \res{x}(\pctx{N}[\pnone{x}] \| \pctx{N'}[\gsome{x}{w_1, \ldots, w_n};Q]) \redone \pctx*{N}[\0] \| \pctx*{N'}[\pnone{w_1} \| \ldots \| \pnone{w_n}]
            }
            \and
            \mathsmaller{\rredone{\equiv}}~~
            \inferrule{
                P \equiv P'
                \\
                P' \redone Q'
                \\
                Q' \equiv Q
            }{
                P \redone Q
            }
            \and
            \mathsmaller{\rredone{\nu}}~~
            \inferrule{
                P \redone P'
            }{
                \res{x}(P \| Q) \redone \res{x}(P' \| Q)
            }
            \and
            \mathsmaller{\rredone{\|}}~~
            \inferrule{
                P \redone P'
            }{
                P \| Q \redone P' \| Q
            }
            \and
            \mathsmaller{\rredone{\nd}}~~
            \inferrule{
                P \redone P'
            }{
                P \nd Q \redone P' \nd Q
            }
        \end{mathpar}
    \end{mdframed}
    \caption{
        Eager reduction semantics for \texorpdfstring{\fullclpi}{spi+}.
    }
    \label{f:eagerReductions}
\end{figure}

The eager reduction semantics, denoted $\redone$, is given in \Cref{f:eagerReductions}.
Intuitively, we follow the principles of Rule~\eqref{eq:usualpi}, discussed earlier:
a reduction step simultaneously expresses (i)~the intended interaction (say, a synchronization) and (ii)~the transformation of the involved contexts so as to express commitment.
To this end, we rely on \emph{ND-contexts}, denoted $\,\pctx{N}, \pctx{M}, \ldots $, and their \emph{commitment}, denoted $\,\D{\pctx{N}}, \D{\pctx{M}}, \ldots $, respectively.

\begin{definition}
    \label{d:ndctx}
    We define \emph{ND-contexts} ($\,\pctx{N},\pctx{M}$) as follows:
    \[
        \pctx{N},\pctx{M} ::= \mhole \sepr \pctx{N} \| P \sepr \res{x}(\pctx{N} \| P) \sepr \pctx{N} \nd P
    \]
    The process obtained by replacing $\mhole$ in $\pctx{N}$ with $P$ is denoted $\pctx{N}[P]$.
    We refer to ND-contexts that do not use the clause `\,$\pctx{N} \nd P$' as \emph{D-contexts}, denoted $\pctx{C},\pctx{D}$.
\end{definition}


This semantics implements the expected commitment of non-deterministic choices by transforming ND-contexts to D-contexts as follows:

\begin{definition}
    \label{d:ncoll}
    The \emph{commitment} of an ND-context $\pctx{N}$, denoted $\D{\pctx{N}}$, is defined as follows:
    \begin{align*}
        \D{\mhole} &:= \mhole
        & \D{\pctx{N} \|  P} &:= \D{\pctx{N}} \| P
        & \D{\res{x}(\pctx{N} \| P)} &:= \res{x}(\D{\pctx{N}} \| P)
        & \D{\pctx{N} \nd P} &:= \D{\pctx{N}}
    \end{align*}
\end{definition}



Barring non-deterministic choice, the reduction rules in \Cref{f:eagerReductions} arise as directed interpretations of proof transformations in the  underlying linear logic. 
We follow Caires and Pfenning~\cite{CairesP10} and Wadler~\cite{DBLP:conf/icfp/Wadler12} in interpreting cut-elimination in linear logic as synchronization in $\clpi$. 
As such, reduction rules are standard but extended with non-determinism: the synchronizing subprocesses appear under ND-contexts (Def.~\labelcref{d:ndctx}) before reductions, and under collapsed ND-contexts (Def.~\labelcref{d:ncoll}) after reductions.
For example, in Rule~$\rredone{{\tensor}{\parr}}$, the send and receive appear under ND-contexts $\pctx{N}$ and $\pctx{N'}$, respectively.
After reduction, these contexts collapse to $\D{\pctx{N}}$ and $\D{\pctx{N'}}$, respectively; notice how the scope of $\D{\pctx{N}}$ extends to the entire process after the reduction, to avoid freeing names that were bound before by $\pctx{N}$.

\begin{example}
    We revisit Example~\labelcref{x:lazy} now under the eager semantics.
    We can express $\res{s} ( \sff{Movies}_s \| \sff{Eve}_s )$ using ND-contexts.
    There is only one way to do so for $\sff{Movies}_s$, since it is deterministic:
    \begin{mathpar}
        \sff{Movies}_s = \pctx{N}[ \gsel{s} \{ \sff{buy} : \sff{MoviesBuy}_s , \sff{peek} : \sff{MoviesPeek}_s \} ]
        \and
        \text{where}
        \and
        \pctx{N} := \mhole.
    \end{mathpar}
    There are three ways to do so for $\sff{Eve}_s$, since there are three non-deterministic branches; for example:
    \begin{mathpar}
        \sff{Eve}_s = \pctx{N'}[ \psel{s}{\sff{buy}} ; \psel{s}{\sff{cash}} ; \sff{EveBuyCash}_s ]
        \and
        \text{where}
        \and
        \pctx{N'} := \psel{s}{\sff{buy}} ; \psel{s}{\sff{card}} ; \sff{EveBuyCard}_s \nd \mhole \nd \psel{s}{\sff{peek}} ; \sff{EvePeek}_s.
    \end{mathpar}
    With $\D{\pctx{N}} = \D{\pctx{N'}} = \mhole$, we have, e.g., $        \res{s} ( \sff{Movies}_s \| \sff{Eve}_s ) \redone \res{s} ( \sff{MoviesBuy}_s \| \psel{s}{\sff{cash}} ; \sff{EveBuyCash}_s )
    $.
    (This is one of three possible reductions.)
    Note: the first reduction immediately determines the payment method.
\end{example}


\subsection{Resource Control for \texorpdfstring{\fullclpi}{spi+} via Session Types}
\label{ss:piTypeSys}

We define a session type system for \fullclpi\kern-.7ex, following `propositions-as-sessions'~\cite{CairesP10,DBLP:conf/icfp/Wadler12}.
As already mentioned, in a session type system,
resources are names that perform protocols:
the \emph{type assignment} $x:A$ says that $x$ should conform to the protocol specified by the session type $A$.
We give the syntax of types:
\begin{align*}
    A,B &::= \1 \sepr \bot \sepr A \tensor B \sepr A \parr B \sepr {\oplus}\{i:A\}_{i \in I} \sepr {?}A \sepr {!}A
    \sepr {\with}\{i:A\}_{i \in I} \sepr {\with}A \sepr {\oplus}A
\end{align*}
The units $\1$ and $\bot$ type closed sessions.
$A \tensor B$ types a name that first outputs a name of type~$A$ and then proceeds as   $B$.
Similarly, $A \parr B$ types a name that inputs a name of type $A$ and then proceeds as~$B$.
Types ${\oplus}\{i:A_i\}_{i \in I}$ and  ${\with}\{i:A_i\}_{i \in I}$ are given to names that can select and offer a labeled choice, respectively.

Type ${?}A$ is assigned to a name that performs a request for a session of type $A$, and type ${!}A$ is assigned to a name that offers a server of type $A$: the modalities `${?}$' and `${!}$' define \emph{non-linear} (i.e., persistent) types.
 Then, ${\with}A$ is the type of a name that \emph{may produce} a behavior of type $A$, or fail; dually, ${\oplus}A$ types a name that \emph{may consume} a behavior of type $A$.

For any type $A$ we denote its \emph{dual} as $\ol{A}$.
Intuitively,  duality of types serves to avoid communication errors: the type at one end of a channel is the dual of the type at the opposite end.
Duality is an involution, defined as follows:
\begin{align*}
    \ol{\1} &= \bot
    & \ol{A \tensor B} &= \ol{A}\parr\ol{B}
    & \ol{{\oplus} \{i:A_i\}_{i\in I}} &= {\with}\{i:\ol{A_i}\}_{i\in I}
    & \ol{{\with}A} &= {\oplus}\ol{A}
    & \ol{{?}A} &= {!}\ol{A}
    \\
    \ol{\bot} &= \1
    & \ol{A \parr B} &= \ol{A}\tensor \ol{B}
    & \ol{{\with}\{i:A_i\}_{i\in I}} &= {\oplus}\{i:\ol{A_i} \}_{i\in I}
        & \ol{{\oplus}A} &= {\with}\ol{A}
        & \ol{{!}A} &= {?}\ol{A}
\end{align*}

Judgments are of the form $P\vdash \Gamma$, where $P$ is a process and $\Gamma$ is a context, a collection of type assignments.
In writing $\Gamma, x:A$, we assume $x \notin \dom{\Gamma}$.
We write $\dom{\Gamma}$ to denote the set of names appearing in $\Gamma$.
We write $\with \Gamma$ to denote that $\forall x:A \in \Gamma.~ \exists A'.~ A = \with  A'$.

\begin{figure}[!t]
    \begin{mdframed} \mysmall
        \begin{mathpar}
            {\ttype[\scriptsize]{cut}}~
            \inferrule{
                P \vdash \Gamma, x{:}A
                \\
                Q \vdash \Delta, x{:}\ol{A}
            }{
                \res{x}(P \| Q) \vdash \Gamma, \Delta
            }
            \and
            \ttype[\scriptsize]{mix}~
            \inferrule{
                P \vdash \Gamma
                \\
                Q \vdash \Delta
            }{
                P \| Q \vdash \Gamma, \Delta
            }
            \and
            \ttype[\scriptsize]{$\nd$}~
            \inferrule{
                P \vdash \Gamma
                \\
                Q \vdash \Gamma
            }{
                P \nd Q \vdash \Gamma
            }
            \and
            \ttype[\scriptsize]{empty}~
            \inferrule{ }{
                \0 \vdash \emptyset
            }
            \and
            \ttype[\scriptsize]{id}~
            \inferrule{ }{
                \pfwd{x}{y} \vdash x{:}A, y{:}\ol{A}
            }
            \and
            \ttype[\scriptsize]{$\1$}~
            \inferrule{ }{
                \pclose{x} \vdash x{:}\1
            }
            \and
            \ttype[\scriptsize]{$\bot$}~
            \inferrule{
                P \vdash \Gamma
            }{
                \gclose{x};P \vdash \Gamma, x{:}\bot
            }
            \and
            \ttype[\scriptsize]{$\tensor$}~
            \inferrule{
                P \vdash \Gamma, y{:}A
                \\
                Q \vdash \Delta, x{:}B
            }{
                \pname{x}{y};(P \| Q) \vdash \Gamma, \Delta, x{:}A \tensor B
            }
            \and
            \ttype[\scriptsize]{$\parr$}~
            \inferrule{
                P \vdash \Gamma, y{:}A, x{:}B
            }{
                \gname{x}{y}; P \vdash \Gamma, x{:}A \parr B
            }
            \and
            \ttype[\scriptsize]{$\oplus$}~
            \inferrule{
                P \vdash \Gamma, x{:}A_j
                \\
                j \in I
            }{
                \psel{x}{j};P \vdash \Gamma, x{:}{\oplus}\{i:A_i\}_{i \in I}
            }
            \and
            \ttype[\scriptsize]{$\with$}~
            \inferrule{
                \forall i \in I.~ P_i \vdash \Gamma, x{:}A_i
            }{
                \gsel{x}\{i:P_i\}_{i \in I} \vdash \Gamma, x{:}{\with}\{i:A_i\}_{i \in I}
            }
            \and
            \ttype[\scriptsize]{${\with}\some$}~
            \inferrule{
                P \vdash \Gamma, x{:}A
            }{
                \psome{x};P \vdash \Gamma, x{:}{\with}A
            }
            \and
            \ttype[\scriptsize]{${\with}\none$}~
            \inferrule{ }{
                \pnone{x} \vdash x{:}{\with}A
            }
            \and
            \ttype[\scriptsize]{${\oplus}\some$}~
            \inferrule{
                P \vdash {\with}\Gamma,  x{:}A
            }{
                \gsome{x}{\dom{\Gamma}};P \vdash {\with}\Gamma,   x{:}{\oplus}A
            }
            \and
            \ttype[\scriptsize]{${?}$}~
            \inferrule{
                P \vdash \Gamma, y{:}A
            }{
                \puname{x}{y};P \vdash \Gamma, x{:}{?}A
            }
            \and
            \ttype[\scriptsize]{${!}$}~
            \inferrule{
                P \vdash {?}\Gamma, y{:}A
            }{
                \guname{x}{y};P \vdash {?}\Gamma, x{:}{!}A
            }
            \and
            \ttype[\scriptsize]{weaken}~
            \inferrule{
                P \vdash \Gamma
            }{
                P \vdash \Gamma, x{:}{?}A
            }
            \and
            \ttype[\scriptsize]{contract}~
            \inferrule{
                P \vdash \Gamma, x{:}{?}A, x'{:}{?}A
            }{
                P\{x/x'\} \vdash \Gamma, x{:}{?}A
            }
        \end{mathpar}
    \end{mdframed}
    \caption{Typing rules for \fullclpi.}
    \label{fig:trulespi}
\end{figure}

\Cref{fig:trulespi} gives the typing rules: they correspond to the rules in Curry-Howard interpretations of classical linear logic as session types (cf.\ Wadler~\cite{DBLP:conf/icfp/Wadler12}), with the rules for ${\with}A$ and ${\oplus}A$ extracted from~\cite{CairesP17}, and the additional Rule~\ttype{$\nd$} for non-confluent non-deterministic choice, which modifies the confluent rule in~\cite{CairesP17}.

We discuss selected rules.
Rule~\ttype{${\with}\some$} types a process with a name whose behavior can be provided, while Rule~\ttype{${\with}\none$} types a name whose behavior cannot.
Rule~\ttype{${\oplus}\some$} types a process with a name $x$ whose   behavior may not be available.
If the behavior is not available, all  the sessions in the process must be canceled; hence, the rule requires all names to be typed under the ${\with}A$ monad.
Rule~\ttype{$\nd$} types our non-deterministic choice operator;  the branches must be typable under the same typing context.
Hence, all branches denote the same sessions, which may be implemented differently.
In context of a synchronization, branches that are kept are able to synchronize, whereas the discarded branches are not; nonetheless, the remaining branches still represent different implementations of the same sessions.

Our type system ensures \emph{session fidelity} (processes correctly follow their ascribed session protocols)
\emph{communication safety} (no communication errors/mismatches occur).
Both properties follow from the fact that typing is consistent across structural congruence and reduction.

We state the main results of this section:
 \fullclpi with $\redone$ satisfies type preservation and deadlock-freedom.

\begin{theorem}[Type Preservation]\label{t:typePresEager}
    If $P \vdash \Gamma$, then both $P \equiv Q$ and $P \redone Q$ imply $Q \vdash \Gamma$.
\end{theorem}


\begin{restatable}[Deadlock-freedom]{theorem}{thmDlfreeOne}\label{t:dlfreeOne}
    If $P \vdash \emptyset$ and $P \not\equiv \0$, then there is $R$ such that $P \redone R$.
\end{restatable}



\section{A Non-deterministic \texorpdfstring{$\lambda$}{Lambda}-calculus with Unrestricted Resources}\label{s:lambda}

We present \lamcoldetsh, a resource  $\lambda$-calculus
with   \nond and  lazy evaluation. Our calculus features  two kinds of resources: linear (to be used exactly once) and unrestricted (usable zero or more times).
It extends the resource $\lambda$-calculus studied in~\cite{DBLP:conf/aplas/HeuvelPNP23}, which does not support unrestricted resources.
In \lamcoldetsh,   \nond is  \col and \emph{implicit}, as it arises from the fetching of terms from {bags} of linear and  unrestricted resources.
(In contrast,   the choice operator `$\nd$' in $\fullclpi$  specifies \nond \emph{explicitly}.)
A mismatch between the number of variable occurrences and the size of the bag induces \emph{failure}.

\subsection{Syntax}

\Cref{f:lambda} gives the syntax of \lamcoldetsh-terms ($M,N,L$) and bags~($A, B$) and contexts~($\lctx{C},\lctx{C}'$).
Variables $x,y,z,\ldots $ have linear and unrestricted occurrences. 
\begin{itemize}
	\item 
Notation $x[\mtt{l}]$ denotes a \emph{linear} occurrence of $x$; we often omit the annotation `$[\mtt{l}]$', and a sequence $\widetilde{x}$ (finite sequence of pairwise distinct $x_i$'s, with length $|\widetilde{x}|$) always involves linear occurrences.
\item Notation~$x[i]$ denotes an \emph{unrestricted} occurrence of $x$, explicitly referencing the $i$-th element of an unrestricted (ordered) bag. We write $\unvar{x}$ whenever the $i$ in $x[i]$ is unimportant.
\end{itemize}

A bag is split into a linear and an unrestricted part:  linear resources in bags cannot be duplicated, but unrestricted resources are always duplicated when consumed. The empty linear bag is denoted $\oneb$.
We use $C_i$ to denote the $i$-th term in the linear bag $C$; also, $\size{C}$ denotes the number of elements in $C$.
To ease readability, we often write, e.g., $\bag{N_1, N_2}$
as a shorthand notation for
$\bag{N_1} \cdot \bag{N_2}$. The empty unrestricted bag is denoted $\unvar{\oneb}$.
Notation $U_i$ denotes the singleton bag at the $i$-th position in $U$; if there is no $i$-th position in $U$, then $U_i$ defaults to $\unvar{\oneb}$.
We use `$\bagsep$' to combine a linear and an unrestricted bag, and unrestricted bags are joined via the non-commutative `$\concat$'.

\begin{figure}[t]
    \begin{mdframed} \mysmall
        \begin{align*}
            M,N,L ::=~
            & x[*] & \text{variable}
            & \quad
            {}\sepr M[\widetilde{x} \leftarrow x] & \text{sharing}
            \\
            \sepr~
            & \lambda x.M & \text{abstraction}
            & \quad
            {}\sepr M \esubst{B}{x} & \text{intermediate substitution}
            \\
            \sepr~
            & (M\ B) & \text{application}
            & \quad
            {}\sepr M \linexsub{C/x_1,\ldots,x_k} & \text{linear substitution}
            \\
            \sepr~
            & \fail^{\tilde{x}} & \text{failure}
            & \quad
            {}\sepr M \unexsub{U/\lunvar{x}} & \text{unrestricted substitution}
            \\
            [*] ::=~
            & [\mtt{l}]
            \sepr
            [i] ~~ i \in \mathbb{N} & \text{annotations}
            & \quad\hphantom{{}\sepr{}}
            A,B ::=~
            C \bagsep U & \text{bag}
            \\
            U,V ::=~
            & \unvar{\oneb} \sepr {\unvar{\bag{M}}} \sepr {U \concat V} & \text{unrestricted bag}
            & \quad\hphantom{{}\sepr{}}
            C,D ::= \oneb \sepr \bag{M} \cdot\, C & \text{linear bag}
            \\
            \lctx{C} ::=~
            & \hole \sepr (\lctx{C}\ B) \sepr \lctx{C} \linexsub{C/\widetilde{x}} \sepr \lctx{C} \unexsub{U/\lunvar{x}} \sepr \lctx{C}[\widetilde{x} \leftarrow x]
            \span\span
            & \text{context}
        \end{align*}
    \end{mdframed}
    \caption{Syntax of $\lamcoldetsh$.}
    \label{f:lambda}
\end{figure}

An abstraction $\lambda x. M$ binds
occurrences  of  $x$ in $M$.
Application $(M\ C)$ is as usual.
The term $M \esubst{B}{x}$ is an intermediate substitution, which involves a bag $B$ with linear and unrestricted parts.
To distinguish between linear and unrestricted occurrences of variables, we have two forms of explicit substitution:
\begin{itemize}
    \item A \emph{linear} substitution 
    of a bag $C$ for $\widetilde{x}$ in  $M$ is denoted $M \linexsub{C /  \widetilde{x}}$.
    We require $\size{C} = |\widetilde{x}|$ and for each $x_i \in \widetilde{x}$: (i)~$x_i$ occurs in $M$; (ii)~$x_i$ is not a sharing variable; (iii)~$x_i$ cannot occur in another explicit substitution in $M$. 
    \item An \emph{unrestricted} substitution, denoted $M \unexsub{U / {x}}$,  concerns the substitution of a  bag $U$ for a  variable $x^!$ in  $M$, with the assumption that $\unvar{x}$ does not appear in another unrestricted substitution in $M$.
\end{itemize}

The \emph{sharing} construct  $M\sharing{x_1,\ldots, x_n}{x}$,
expresses that
$x$  may be used in $M$ under ``aliases'' $x_1,\ldots, x_n$. Hence, it atomizes $n$ occurrences of   $x$   in~$M$, via an explicit pointer to $n$ variables.
 In  $M\sharing{\widetilde{x}}{x}$, we say that $\widetilde{x}$ are the \emph{shared variables}  and that $x$  is the \emph{sharing variable}.
We require for each $x_i \in \widetilde{x}$: (i)~$x_i$ occurs exactly once in $M$; (ii)~$x_i$ is not a sharing variable.
The sequence $\widetilde{x}$ can be empty:
$M\sharing{}{x}$ means that $x$ does not share any variables in $M$.
Sharing binds the shared variables in the term.
This way, e.g.,
the $\lambda$-term $\lambda x.(x\ x)$ is expressed in $\lamcoldetsh$ as $\lambda x. (x_1 \bag{x_2}\sharing{x_1,x_2}{x})$, where $\bag{x_2}$ is a bag containing~$x_2$.

The term $\fail^{\tilde{x}}$ denotes failure; the variables in $\widetilde{x}$ are ``dangling'' resources, which cannot be accounted for after failure.
We write $\lfv{M}$ to denote the free variables of $M$.
Term $M$ is \emph{closed} if $\lfv{M} = \emptyset$.

\begin{figure}[t]
    \begin{mdframed} \mysmall
        \begin{mathpar}
            \inferrule[$\redlab{RS{:}Beta}$]{ }{
                (\lambda x . M)\ B  \red M \esubst{ B }{ x }
            }
            \and
            \inferrule[$\redlab{RS{:}Ex \dash Sub}$]{
                \size{C} = |\widetilde{x}|
                \\
                M \not= \fail^{\tilde{y}}
            }{
                (M\sharing{\widetilde{x}}{x}) \esubst{ C \bagsep U }{ x } \red  M \linexsub{C  /  \widetilde{x}} \unexsub{U / \lunvar{x} }
            }
            \and
            \inferrule[$\redlab{RS{:}Fetch^{\ell}}$]{
                \headf{M} =  {x}_j
                \\
                0 < i \leq \size{C}
            }{
                M \linexsub{C /  \widetilde{x}, x_j} \red  (M \headlin{ C_i / x_j })  \linexsub{(C \setminus C_i ) /  \widetilde{x}  }
            }
            \and
            \inferrule[$\redlab{RS{:}Fail^{\ell}}$]{
                \size{C} \neq |\widetilde{x}|
                \\
                \widetilde{y} = (\llfv{M} \setminus \{  \widetilde{x}\} ) \cup \llfv{C}
            }{
                (M\sharing{\widetilde{x}}{x}) \esubst{C \bagsep U}{ x }  \red  \fail^{\widetilde{y}}
            }
            \and
            \inferrule[$\redlab{RS{:} Fetch^!}$]{
                \headf{M} = {x}[i]
                \\
                U_i = \unvar{\bag{N}}
            }{
                M \unexsub{U / \lunvar{x}} \red  M \headlin{ N /{x}[i] }\unexsub{U / \lunvar{x}}
            }
            \and
            \inferrule[$\redlab{RS{:}Fail^!}$]{
                \headf{M} = {x}[i]
                \\
                U_i = \unvar{\oneb}
            }{
                M \unexsub{U / \lunvar{x} } \red M \headlin{ \fail^{\emptyset} /{x}[i] } \unexsub{U / \lunvar{x} }
            }
            \and
            \inferrule[$\redlab{RS{:}Cons_1}$]{
                \widetilde{y} = \llfv{C}
            }{
                \fail^{\widetilde{x}}\ (C \bagsep U) \red  \fail^{\widetilde{x} \ \widetilde{y}}
            }
            \and
            \inferrule[$\redlab{RS{:}Cons_2}$]{
                \size{C} =   |  {\widetilde{x}} |
                \\
                \widetilde{z} = \llfv{C}
            }{
                (\fail^{ {\widetilde{x}} \cup \widetilde{y}} \sharing{\widetilde{x}}{x}) \esubst{ C \bagsep U }{ x }  \red  \fail^{\widetilde{y} \cup \widetilde{z}}
            }
            \and
            \inferrule[$\redlab{RS{:}Cons_3}$]{
                \widetilde{z} = \llfv{C}
            }{
                \fail^{\widetilde{y}\cup \widetilde{x}} \linexsub{C /  \widetilde{x}} \red  \fail^{\widetilde{y} \cup \widetilde{z}}
            }
            \and
            \inferrule[$\redlab{RS{:}Cons_4}$]{ }{
                \fail^{\widetilde{y}} \unexsub{U / \lunvar{x}}  \red  \fail^{\widetilde{y}}
            }
            \and
            \inferrule[$\redlab{RS:TCont}$]{
                M \red    N
            }{
                \lctx{C}[M] \red   \lctx{C}[N]
            }
        \end{mathpar}
        
                    where $\headf{M}$ is defined as follows:
        \begin{align*}
            \headf{ x } &= x
            &
            \head{x[i]} &= x[i] 
            &
            \headf{ \lambda x.M } &= \lambda x.M
            &
            \headf{ (M\ C) } &= \headf{M}
            \\
            \headf{ \fail^{\widetilde{x}} } &= \fail^{\widetilde{x}}
            &
            \headf{ M \esubst{ C }{ x } } &= M \esubst{ C }{ x }
            &
            \headf{ M \linexsub{C / \widetilde{x}} } &= \headf{M}
            &            
             \head{M \unexsub{U/x}} &= \head{M}\\
            \\
            \headf{ M\sharing{\widetilde{x}}{x} } = \begin{cases}
                x & \text{$\headf{ M } = y$ and $y \in  {\widetilde{x}}$}
                \\
                \headf{ M } & \text{otherwise}
            \end{cases}
            \span\span\span\span
        \end{align*}
        
    \end{mdframed}

    \caption{Reduction rules for \texorpdfstring{\lamcoldetsh}{lambda}.}
    \label{fig:reduc_interm}\label{f:lambda_red}
\end{figure}

\subsection{Reduction Semantics}

\Cref{f:lambda_red} gives the reduction semantics  for $\lamcoldetsh$, denoted $\red$.
 Rule~$\redlab{RS:Beta}$ induces an intermediate explicit substitution. Rule~$\redlab{RS:Ex{\dash}Sub}$  reduces an intermediate substitution to an explicit substitution that will manage the two-component format of bags.
An explicit substitution $(M\sharing{\widetilde{x}}{x}) \esubst{ C \bagsep U }{ x }$ reduces to a term in which the linear and unrestricted parts of the bag are separated into their explicit substitutions $M\linexsub{C  /  \widetilde{x}} \unexsub{U / \lunvar{x} }$. This only occurs when the size of the bag equals the number of shared variables.
 The fetching of linear/unrestricted resources from their corresponding bags is done by the appropriated fetch rules~$\redlab{RS:Fetch^\ell}$ or $\redlab{RS:Fetch^!}$.  In case of a mismatch, the term evolves into failure via Rule~$\redlab{RS:Fail^\ell}$.

 An explicit substitution $M \linexsub{C/\widetilde{x}}$, where the head variable of $M$ is $x_j \in \widetilde{x}$, reduces via Rule~$\redlab{RS:Fetch^\ell}$. The rule extracts a $C_i$ from $C$ (for some $0 < i \leq \size{C}$) and substitutes it for $x_j$ in $M$; this is how fetching induces a non-deterministic choice between $\size{C}$ possible reductions.
The reduction of an unrestricted substitution $M \unexsub{U/\lunvar{x}}$, where the head variable of $M$ is $x[i]$, depends on $U_i$:
\begin{itemize}
    \item
        If $U_i = \unvar{\bag{N}}$, then the term reduces via Rule~$\redlab{R:Fetch^!}$ by substituting the head occurrence of $x[i]$ in $M$ with $N$, denoted $M\headlin{N/x[i]}$; note that $U_i$ remains available after this reduction.

    \item
        If $U_i = \unvar{\oneb}$, the head variable is instead substituted with failure via Rule~$\redlab{R:Fail^!}$.
\end{itemize}
  Rules~$\redlab{RS:Cons_j}$ for $j \in \{1,2,3\}$ consume terms when they meet failure.
Finally, Rule~$\redlab{RS:TCont}$ closes reduction under contexts.
The following example illustrates reduction.

\begin{example}\label{ex:syntax}\label{ex:lambdaRed} 
    Consider the term $M_0 = ( \lambda x. x_1 \bag{x_2 \bag{x_3\ \oneb} } \sharing{ \widetilde{x} }{x} )\ \bag{\fail^{\emptyset} , y ,I\,}$,
    where $I = \lambda x. (x_1 \sharing{x_1}{x})$ and $\widetilde{x} = x_1 , x_2 ,x_3$. Here we simplify the notation of $M_0$  by omitting the empty unrestricted bag, i.e., we write $M\bag{M'} $ to denote $M\bag{M'}\bagsep \oneb^!$.
    First, $M_0$ evolves into an intermediate substitution~\eqref{eq:lin_cons_sub1}.
    The bag can provide for all shared variables, so it then evolves into an explicit substitution~\eqref{eq:lin_cons_sub2}:
    \begin{align}
        M_0
        &\red  (x_1 \!\bag{\!x_2 \!\bag{\!x_3\ \oneb\!} \!}  \sharing{\widetilde{x}}{x}) \esubst{ \bag{\fail^{\emptyset} , y , I}   }{x}\unexsub{\oneb^! / \lunvar{x} }
        \label{eq:lin_cons_sub1}
        \\
        &\red
        (x_1 \!\bag{\!x_2 \!\bag{\!x_3\ \oneb\!}  \!}) \linexsub{\! \bag{\!\fail^{\emptyset} , y , I\!} \!/ \widetilde{x} }\unexsub{\oneb^! / \lunvar{x} } = M
        \label{eq:lin_cons_sub2}
    \end{align}
    Since $ \headf{M} = x_1$, one of the three elements of the bag will be substituted.
    $M$ represents a non-deterministic choice between the following three reductions:
    \begin{align*}
        \mathbin{\rotatebox[origin=r]{30}{$\red$}} &~ (\fail^{\emptyset} \bag{x_2 \bag{x_3\ \oneb}  }) \linexsub{  \bag{y, I} /  x_2,x_3  }\unexsub{\oneb^! / \lunvar{x} }
        = N_1
        \\[-5pt]
        M~
        \red &~ (y \bag{x_2 \bag{x_3\ \oneb}  }) \linexsub{ \bag{\fail^{\emptyset} , I} /  x_2,x_3  }\unexsub{\oneb^! / \lunvar{x} }
        = N_2
        \\[-3pt]
        \mathbin{\rotatebox[origin=r]{-30}{$\red$}} &~ (I \bag{x_2 \bag{x_3\ \oneb}}) \linexsub{ \bag{\fail^{\emptyset} , y}  /  x_2,x_3  }\unexsub{\oneb^! / \lunvar{x} }
        = N_3
    \end{align*}
   There are no rules for garbage collection; therefore, the normal forms maintain the explicit substitution for empty unrestricted bags.
\end{example}

\subsection{Resource Control for \texorpdfstring{\lamcoldetsh}{Lambda} via Intersection Types}\label{sec:lamTypes}

Our type system for   $\lamcoldetsh$ is based on non-idempotent intersection types which, 
As in~\cite{PaganiR10,DBLP:conf/birthday/BoudolL00}, intersection types account for available resources in bags, which are unordered and have all the same type.
Because we admit the term $\fail^{\tilde x}$ as typable, we say that our system enforces \emph{well-formedness} rather than \emph{well-typedness}.
As we will see, well-typed terms form the sub-class of well-formed terms that does not include $\fail^{\tilde x}$. 

Strict types ($\sigma, \tau, \delta$) and multiset types ($\pi, \zeta$) are defined as follows:
\begin{align*}
    \sigma, \tau, \delta ::=~ &
    \unit \sepr \arrt{ (\pi , \eta) }{\sigma}
  &  \pi, \zeta ::=
    \bigwedge_{i \in I} \sigma_i \sepr \omega
\\
    \eta, \epsilon  ::=~ &
    \sigma  \sepr \epsilon \concat \eta
    \qquad \text{list}
    &
    ( \pi , \eta)
    \qquad \text{tuple}
\end{align*}
\noindent
Given a non-empty $I$, multiset types $\bigwedge_{i\in I}\sigma_i$ are given to bags of size $|I|$.
This operator is associative, commutative,  and non-idempotent (i.e., $\sigma\wedge\sigma\neq \sigma$), with identity $\omega$.
Notation $\sigma^k$ stands for $\sigma \wedge \cdots \wedge \sigma$ ($k$ times, if $k>0$) or $\omega$ (if $k=0$).

The list type $\epsilon\concat \eta$ types the concatenation of unrestricted bags.
It can be recursively unfolded into a finite composition of strict types $\sigma_1\concat \ldots\concat \sigma_n$, for some $n\geq 1$, with length $n$ and $\sigma_i$ its $i$-th strict type ($1\leq i\leq n$).
We  write $\unvar{x}:\eta $ to denote for $x[1]:\eta_1 , \ldots , x[k]:\eta_k$ where $\eta$ has length $k$.
The tuple type $(\pi,\eta)$ types concatenation of a linear bag of type $\pi$ with an unrestricted bag of type $\eta$. Finally strict types are amended to allow for unrestricted functional types which go from tuple types to strict types \arrt{ (\pi , \eta) }{\sigma} rather then multiset types to strict types.

\begin{definition}[$\eta \relunbag \epsilon$]\label{not:ltypes}
    Let $\epsilon$ and $\eta$ be two list types, with the length of $\epsilon$ greater or equal to that of $\eta$.
    We say that $\epsilon$ \emph{embraces} $\eta$, denoted $\eta \relunbag \epsilon$, whenever there exist $  \epsilon'$ and $ \epsilon''$   such that: i) $ \epsilon = \epsilon' \concat \epsilon''$; ii)  the size of $\epsilon' $ is that of $\eta$; iii)  for all $i$, $\epsilon'_i = \eta_i $.

\end{definition}

We separate contexts into two parts: linear ($\Gamma, \Delta,\ldots$) and unrestricted ($\Theta,\Upsilon,\ldots$):
\begin{align*}
    \Gamma,\Delta &::= \dash \sep \Gamma , x:\pi  \sepr \Gamma, x:\sigma
    &
    \Theta,\Upsilon &::= \dash \sepr \Theta, x^!:\eta
\end{align*}
where $\dash$ denotes the empty context. Both linear and unrestricted occurrences of variables may occur at most once in a context.
Judgments have the form $\Theta;\Gamma\wfdash M:\tau$.
We write $\wfdash M:\tau$ to denote $\dash;\dash\wfdash M:\tau$.

We write $\dom{\Gamma}$ for the set of variables in~$\Gamma$.
For $\Gamma, x:\pi$, we assume $x \not \in \dom{\Gamma}$.
To avoid ambiguities, we write $x:\sigma^1$  to denote that the assignment involves a multiset type, rather than a strict~type.
Given $\Gamma$, its \emph{core context} $\core{\Gamma}$ concerns variables with types different from $\omega$; it is defined as
$\core{\Gamma} = \{ x:\pi \in \Gamma \,|\, \pi \not = \omega\}$.

\begin{definition}[Well-formedness in $\lamcoldetsh$]
    A $\lamcoldetsh$-term $M$ is \emph{well-formed} if there exists a context $\Theta$ and $\Gamma$ and a  type  $\tau$ such that the rules in \Cref{fig:wfsh_rulesunres} entail $\Theta ; \Gamma \wfdash  M : \tau $.
\end{definition}

\begin{figure}[!t]
    \begin{mdframed} \mysmall
        \begin{mathpar}
            \inferrule[$\redlab{FS{:}var^{\ell}}$]{ }{
                \Theta ;  {x}: \sigma \wfdash  {x} : \sigma
            }
            \and
            \inferrule[$\redlab{FS{:}var^!}$]{
                \Theta , x^!: \eta;  {x}: \eta_i , \Delta \wfdash  {x} : \sigma
            }{
                \Theta ,  x^!: \eta; \Delta \wfdash {x}[i] : \sigma
            }
            \and
            \inferrule[$\redlab{FS{:}\oneb^{\ell}}$]{ }{
                \Theta ; \dash \wfdash \oneb : \omega
            }
            \and
            \inferrule[$\redlab{FS{:}bag^{\ell}}$]{
                \Theta ; \Gamma \wfdash M : \sigma
                \\
                \Theta ; \Delta \wfdash C : \sigma^k
            }{
                \Theta ; \Gamma , \Delta \wfdash \bag{M}\cdot C:\sigma^{k+1}
            }
            \and
            \inferrule[$\redlab{FS{:}\oneb^!}$]{ }{
                \Theta ;  \dash  \wfdash \unvar{\oneb} : \sigma
            }
            \and
            \inferrule[$\redlab{FS{:} bag^{!}}$]{
                \Theta ; \dash \wfdash U : \epsilon
                \\
                \Theta ; \dash \wfdash V : \eta
            }{
                \Theta ; \dash  \wfdash U \concat V :\epsilon \concat \eta
            }
            \and
            \inferrule[$\redlab{FS{:}bag}$]{
                \Theta ; \Gamma\wfdash C : \sigma^k
                \\
                \Theta ;\dash \wfdash  U : \eta
            }{
                \Theta ; \Gamma \wfdash C \bagsep U : (\sigma^{k} , \eta  )
            }
            \and
            \inferrule[$\redlab{FS{:}fail}$]{
                \dom{\core{\Gamma}} = \widetilde{x}
            }{
                \Theta ; \core{\Gamma} \wfdash  \fail^{\widetilde{x}} : \tau
            }
            \and
            \inferrule[$\redlab{FS{:}weak}$]{
                \Theta ; \Gamma  \wfdash M : \tau
            }{
                \Theta ; \Gamma ,  {x}: \omega \wfdash M\sharing{}{x}: \tau
            }
            \and
            \inferrule[$\redlab{FS{:}shar}$]{
                \Theta ;  \Gamma ,  {x}_1: \sigma, \cdots,  {x}_k: \sigma \wfdash M : \tau
                \\
                {x}\notin \dom{\Gamma}
                \\
                k \not = 0
            }{
                \Theta ;  \Gamma ,  {x}: \sigma^{k} \wfdash M\sharing{{x}_1 , \ldots ,  {x}_k}{x}  : \tau
            }
            \and
            \inferrule[$\redlab{FS{:}abs\dash sh}$]{
                \Theta , x^!:\eta ; \Gamma ,  {x}: \sigma^k \wfdash M\sharing{\widetilde{x}}{x} : \tau
                \\
                {x} \notin \dom{\Gamma}
            }{
                \Theta ; \Gamma \wfdash \lambda x . (M[ {\widetilde{x}} \leftarrow  {x}])  : (\sigma^k, \eta )  \rightarrow \tau
            }
            \and
            \inferrule[$\redlab{FS{:}app}$]{
                \Theta ;\Gamma \wfdash M : (\sigma^{j} , \eta ) \rightarrow \tau
                \\
                \Theta ;\Delta \wfdash B : (\sigma^{k} , \epsilon )
                \\
                \eta \relunbag \epsilon
            }{
                \Theta ; \Gamma , \Delta \wfdash (M\ B) : \tau
            }
\and
            \inferrule[$\redlab{FS{:}Esub^!}$]{
                \Theta , x^! {:} \eta; \Gamma  \wfdash M : \tau
                \\
                \Theta ; \dash \wfdash U : \epsilon
                \\
                \eta \relunbag \epsilon
            }{
                \Theta ; \Gamma \wfdash M \unexsub{U / {x}}  : \tau
            }
            \and
            \inferrule[$\redlab{FS{:}Esub}$]{
                \Theta , x^! : \eta ; \Gamma ,  {x}: \sigma^{j} \wfdash M[ {\widetilde{x}} \leftarrow  {x}] : \tau
                \\
                \Theta ; \Delta \wfdash B : (\sigma^{k} , \epsilon )
                \\
                \eta \relunbag \epsilon
            }{
                \Theta ; \Gamma , \Delta \wfdash (M[ {\widetilde{x}} \leftarrow  {x}])\esubst{ B }{ x }  : \tau
            }
            \and
            \inferrule[$\redlab{FS{:}Esub^{\ell}}$]{
                \Theta ; \Gamma  ,  x_1:\sigma, \cdots , x_k:\sigma \wfdash M : \tau
                \\
                \Theta ; \Delta \wfdash C : \sigma^k
            }{
                \Theta ; \Gamma , \Delta \wfdash M \linexsub{C /  x_1, \cdots , x_k} : \tau
            }
        \end{mathpar}
    \end{mdframed}
    \caption{Well-Formedness Rules for $\lamcoldetsh$.}
    \label{fig:wfsh_rulesunres}
\end{figure}

\noindent
In \Cref{fig:wfsh_rulesunres},
Rule~$\redlab{FS:var^{\ell}}$ types variables.
Rule~$\redlab{FS:\oneb^{\ell}}$ types the empty bag with $\omega$.
Rule~$\redlab{FS:bag^{\ell}}$ types the concatenation of bags.
Rule~$\redlab{FS:\fail}$ types the term $\fail^{\widetilde{x}}$ with a strict type $\tau$, provided that the domain of the core context coincides with $\widetilde{x}$ (i.e., no  variable in $\widetilde{x}$ is typed with $\omega$).
Rule~$\redlab{FS:weak}$ types $M\sharing{}{x}$ by weakening the context with $x:\omega$.
Rule~$\redlab{FS:shar}$ types $M\sharing{\widetilde{x}}{x}$ with $\tau$, provided that there are assignments to the shared variables in~$\widetilde{x}$.

Rule~$\redlab{FS:abs{\dash}sh}$ types an abstraction $\lambda x.( M\sharing{\widetilde{x}}{x})$ with  $\sigma^k\to \tau$, provided that   $M\sharing{\widetilde{x}}{x}:\tau$ can be entailed from an assignment  $x:\sigma^k$.
Rule~\redlab{FS:app}  types $(M\ C)$, provided that $M$ has type $\sigma^j \to \tau$ and  $C$ has type $\sigma^k$. Note that, unlike usual intersection type systems, $j$ and $k$ may differ.
Rule~$\redlab{FS:Esub}$ types the intermediate substitution of a bag $C$ of type $\sigma^k$, provided that $x$ has type $\sigma^j$; again, $j$ and $k$ may differ.
Rule~$\redlab{FS:Esub^{\ell}}$  types  $M\linexsub{C/\widetilde{x}}$ as long as $C$ has type $\sigma^{|\widetilde{x}|}$,  and  each $x_i \in \widetilde{x}$ is of type $\sigma$.

Well-formed terms satisfy subject reduction (SR):

\begin{theorem}[SR in $\lamcoldetsh$]\label{t:lamSR}
    If $\Theta ; \Gamma \wfdash M:\tau$ and $M \red M'$ then $\Theta ; \Gamma \wfdash M' :\tau$.
\end{theorem}

From our system for well-formedness we can extract a system for \emph{well-typed} terms, which do not include $\fail^{\tilde x}$.
Judgments for well-typedness are denoted $ \Gamma \wtdash M:\tau$, with rules adapted from 
\Cref{fig:wfsh_rulesunres}
(the rule name prefix \texttt{FS} is replaced with \texttt{TS}), with the following modifications:
(i)~There is no rule~$\redlab{TS{:}fail}$; (ii)~Rules~$\redlab{TS{:}app}$ and $\redlab{TS{:}Esub}$ do not allow a mismatch between variables and resources.
This way, e.g.,
\[
    \inferrule[$\redlab{TS{:}Esub}$]{
        \Theta , x^! : \eta ; \Gamma ,  {x}: \sigma^{j} \wtdash M[ {\widetilde{x}} \leftarrow  {x}] : \tau
        \\
        \Theta ; \Delta \wtdash B : (\sigma^{k} , \eta )
    }{
        \Theta ; \Gamma , \Delta \wtdash (M[ {\widetilde{x}} \leftarrow  {x}])\esubst{ B }{ x }  : \tau}
\]
For the sake of completeness, the full set of rules is in \Cref{fig:wtsh_rulesunres}. 
Well-typed terms are also well-formed, and thus satisfy SR.
Moreover, 
well-typed terms
also satisfy subject expansion (SE).

\begin{theorem}[SE in $\lamcoldetsh$]\label{t:lamSE}
If $\Theta ; \Gamma \wtdash M':\tau$ and $M \red M'$ then $\Theta ; \Gamma \wtdash M :\tau$.
\end{theorem}

\begin{figure}[h!]
\begin{mdframed}
\small
\begin{mathpar}
    \inferrule[$\redlab{TS{:}var^{\ell}}$]{ }{
        \Theta;  {x}: \sigma \wtdash  {x} : \sigma
    }
    \and
    \inferrule[$\redlab{TS{:}var^!}$]{
        \Theta , x^!: \eta;  {x}: \eta_i , \Delta \wtdash  {x} : \sigma
    }{
        \Theta ,  x^!: \eta; \Delta \wtdash {x}[i] : \sigma
    }
    \and
    \inferrule[$\redlab{TS{:}\oneb^{\ell}}$]{ }{
        \Theta ; \dash \wtdash \oneb : \omega
    }
    \and
    \inferrule[$\redlab{TS\!:\!weak}$]{
        \Theta ; \Gamma  \wtdash M : \tau
    }{
        \Theta ; \Gamma ,  {x}: \omega \wtdash M[\leftarrow  {x}]: \tau
    }
    \and
    \inferrule[$\redlab{TS{:}abs\dash sh}$]{
        \Theta , x^!:\eta ; \Gamma ,  {x}: \sigma^k \wtdash M[ {\widetilde{x}} \leftarrow  {x}] : \tau
        \\
        {x} \notin \dom{\Gamma}
    }{
        \Theta ; \Gamma \wtdash \lambda x . (M[ {\widetilde{x}} \leftarrow  {x}])  : (\sigma^k, \eta )  \rightarrow \tau
    }
    \and
    \inferrule[$\redlab{TS{:}bag^{\ell}}$]{
        \Theta ; \Gamma \wtdash M : \sigma
        \\
        \Theta ; \Delta \wtdash C : \sigma^k
    }{
        \Theta ; \Gamma , \Delta \wtdash \bag{M}\cdot C:\sigma^{k+1}
    }
    \and
    \inferrule[$\redlab{TS{:}app}$]{
        \Theta ;\Gamma \wtdash M : (\sigma^{j} , \eta ) \rightarrow \tau
        \\
        \Theta ;\Delta \wtdash B : (\sigma^{j} , \eta )
    }{
        \Theta ; \Gamma , \Delta \wtdash M\ B : \tau
    }
    \and
    \inferrule[$\redlab{TS{:} bag^{!}}$]{
        \Theta ; \dash \wtdash U : \epsilon
        \\
        \Theta ; \dash \wtdash V : \eta
    }{
        \Theta ; \dash  \wtdash U \concat V :\epsilon \concat \eta
    }
    \and
    \inferrule[$\redlab{TS{:}shar}$]{
        \Theta ;  \Gamma ,  {x}_1: \sigma, \dots,  {x}_k: \sigma \wtdash M : \tau
        \\
        {x}\notin \dom{\Gamma}
        \\
        k \not = 0
    }{
        \Theta ;  \Gamma ,  {x}: \sigma^{k} \wtdash M [  {x}_1 , \dots ,  {x}_k \leftarrow  {x} ]  : \tau
    }
    \and
    \inferrule[$\redlab{TS{:}bag}$]{
        \Theta ; \Gamma\wtdash C : \sigma^k
        \\
        \Theta ;\dash \wtdash  U : \eta
    }{
        \Theta ; \Gamma \wtdash C \bagsep U : (\sigma^{k} , \eta  )
    }
    \and
    \inferrule[$\redlab{TS{:}Esub^{\ell}}$]{
        \Theta ; \Gamma  ,  x_1:\sigma, \dots , x_k:\sigma \wtdash M : \tau ~~  \Theta ; \Delta \wtdash C : \sigma^k
    }{
        \Theta ; \Gamma , \Delta \wtdash M \linexsub{C /  x_1, \dots , x_k} : \tau
    }
    \and
    \inferrule[$\redlab{TS{:}Esub^!}$]{
        \Theta , x^! {:} \eta; \Gamma  \wtdash M : \tau
        \\
        \Theta ; \dash \wtdash U : \eta
    }{
        \Theta ; \Gamma \wtdash M \unexsub{U / \unvar{x}}  : \tau
    }
    \and
    \inferrule[$\redlab{TS{:}Esub}$]{
        \Theta , x^! : \eta ; \Gamma ,  {x}: \sigma^{j} \wtdash M[ {\widetilde{x}} \leftarrow  {x}] : \tau
        \\
        \Theta ; \Delta \wtdash B : (\sigma^{k} , \eta )
    }{
        \Theta ; \Gamma , \Delta \wtdash (M[ {\widetilde{x}} \leftarrow  {x}])\esubst{ B }{ x }  : \tau
    }
\end{mathpar}
\end{mdframed}
\caption{Well-Typed Rules for $\lamcoldetsh$.}
\label{fig:wtsh_rulesunres}
\end{figure}

\section{Translating \texorpdfstring{\lamcoldetsh}{Unrestricted Lambda} into  \texorpdfstring{\fullclpi}{Pi}}
\label{s:transUnres}

Clearly, \fullclpi and \lamcoldetsh are different models.
In particular, 
\lamcoldetsh is a  programming calculus in which implicit non-determinism
implements  fetching of linear and unrestricted resources.
To illustrate the potential of \fullclpi to precisely model non-determinism as found in realistic programs/protocols using an eager approach, 
we follow and extend the approach in \cite{DBLP:conf/aplas/HeuvelPNP23}, and give a translation of  \lamcoldetsh into \fullclpi which 
preserves types (Theorem~\ref{t:preservationencode}) and respects well-known  criteria for dynamic correctness (Definition~\ref{d:encCriteria}).

\subsection{The translation}
Given a \lamcoldetsh-term \(M\), its translation into \fullclpi is denoted \(\piencod{M}_{u}\) and given in \Cref{fig:encoding}.
As in Milner's classical translation:  every variable $x$ in $M$ becomes a name \(x\) in process \(\piencod{M}_{u}\), where
name \(u\) provides the behavior of \(M\).
To handle failures in \lamcoldetsh, \(u\) is a non-deterministically session: the translated term can be available or not, as signaled by prefixes \(\psome{u}\) and \(\pnone{u}\), respectively.
As a result, reductions from \(\piencod{M}_{u}\) include synchronizations that codify $M$'s behavior but also  synchronizations that confirm a session's availability.

\begin{figure}[t]
    \begin{mdframed} \mysmall
        \begin{align*}
            \piencodf{ {x}}_u & = \psome{x}; \pfwd{x}{u} \hspace{1cm}
            \piencodf{{x}[j]}_u  =   \puname{\unvar{x}}{{x_i}}; \psel{x_i}{j}; \pfwd{x_i}{u}
            \\[1mm]
            \piencodf{\lambda x.M}_u & = \psome{u};\gname{u}{x};  \psome{x};\gname{x}{\linvar{x}}; \gname{x}{\unvar{x}};  \gclose{x} ; \piencodf{M}_u
            \\[1mm]
            \piencodf{ M \esubst{ C \bagsep U }{ x} }_u & =  \res{x}( \psome{x}; \gname{x}{\linvar{x}}; \gname{x}{\unvar{x}};  \gclose{x} ;\piencodf{ M}_u \| \piencodf{ C \bagsep U}_x )
            \\[1mm]
            \piencodf{M (C \bagsep U)}_u & = \res{v} (\piencodf{M}_v \| \gsome{v}{u , \llfv{C}};\pname{v}{x}; ( \piencodf{C \bagsep U}_x   \| \pfwd{v}{u}  ) )
            \\[1mm]
            \piencodf{ C \bagsep U }_x & = \gsome{x}{\llfv{C}};  \pname{x}{\linvar{x}}; \big( \piencodf{ C }_{\linvar{x}} \|  \pname{x}{\unvar{x}}; ( \guname{\unvar{x}}{x_i}; \piencodf{ U }_{x_i}  \| \pclose{x} ) \big)
            \\[1mm]
            \piencodf{\bag{M_j} \cdot~ C}_{\linvar{x}}  &=
            \begin{array}{@{}l@{}}
                \gsome{\linvar{x}}{\llfv{C} }; \gname{x}{y_i}; \gsome{\linvar{x}}{y_i, \llfv{C}}; \psome{\linvar{x}}; \\
                \quad \pname{\linvar{x}}{z_i}; ( \gsome{z_i}{\llfv{M_j}};  \piencodf{M_j}_{z_i} \| (\piencodf{(C \setminus M_j)}_{\linvar{x}} \| \pnone{y_i} ))
            \end{array}
            \\[1mm]
            \piencodf{{\oneb}}_{\linvar{x}} & = \gsome{\linvar{x}}{\emptyset};\gname{x}{y_n};  ( \psome{ y_n}; \pclose{y_n}  \| \gsome{\linvar{x}}{\emptyset}; \pnone{\linvar{x}} )
            \\[1mm]
            \span
            \piencodf{\unvar{\oneb}}_{x}  = \pnone{x}
            \qquad\qquad
            \piencodf{\unvar{\bag{N}}}_{x}  =  \piencodf{N}_{x}
            \qquad\qquad
            \piencodf{ U }_{x}  = \gsel{x}\{i:\piencodf{ U_i }_{x} \}_{U_i \in U}
            \\[1mm]
            \piencodfscale{
            M \ltalltriangle \bag{M_1} \cdot \bag{M_2}
                    {} /  x_1, x_2 \rtalltriangle
            }_u    &= \begin{array}{@{}l@{}}
                \res{z_1}( \gsome{z_1}{\llfv{M_{1}}};\piencodf{ M_{1} }_{ {z_1}} \| \res{z_2} ( \gsome{z_2}{\llfv{M_{2}}};\piencodf{ M_{2} }_{ {z_2}} \\
                \qquad \quad {} \| {} {} \bignd_{x_{i_1} \in \{ x_1 , x_2  \}} \bignd_{x_{i_2} \in \{ x_1, x_2 \setminus x_{i_1}  \}} \piencodf{ M }_u \{ z_1 / x_{i_1} \} \{ z_2 / x_{i_2} \} ) \ldots )
            \end{array}
            \\[1mm]
            \piencodf{ M \unexsub{U / \lunvar{x}}  }_u   & =  \res{\unvar{x}} ( \piencodf{ M }_u \|   \guname{\unvar{x}}{x_i}; \piencodf{ U }_{x_i} )
            \\[1mm]
            \piencodf{M[  \leftarrow  {x}]}_u & = \psome{\linvar{x}}; \pname{\linvar{x}}{y_i}; ( \gsome{y_i}{ u , \llfv{M} }; \gclose{ y_{i} } ;\piencodf{M}_u \| \pnone{ \linvar{x} } )
            \\[1mm]
            \piencodf{M[ \widetilde{x} \leftarrow  {x}]}_u & = \begin{array}{@{}l@{}}
                \psome{\linvar{x}}; \pname{\linvar{x}}{y_i}; \big( \gsome{y_i}{ \emptyset }; \gclose{ y_{i} } ; \0 \\
                \quad {} \| \psome{\linvar{x}}; \gsome{\linvar{x}}{u, \llfv{M} \setminus  \widetilde{x} }; \bignd_{x_i \in \widetilde{x}} \gname{\linvar{x}}{{x}_i};\piencodf{M[ (\widetilde{x} \setminus x_i ) \leftarrow  {x}]}_u \big)
            \end{array}
            \\[1mm]
            \piencodf{\fail^{x_1, \ldots, x_k}}_u & = \pnone{ u}  \| \pnone{ x_1} \| \ldots \| \pnone{ x_k}
        \end{align*}
    \end{mdframed}
    \caption{Translation of \texorpdfstring{\lamcoldetsh}{lambda} into \texorpdfstring{\fullclpi}{spi+}.}\label{fig:encoding}
\end{figure}


We discuss \Cref{fig:encoding}, focusing on constructs related to unrestricted resources, not considered in \cite{DBLP:conf/aplas/HeuvelPNP23}.
The translation of an unrestricted variable $x[j]$ first connects to a server along channel $x$ via a request $ \puname{\unvar{x}}{{x_i}}$ followed by a selection on $ \psel{x_i}{j}$.
Process $\piencodf{\lambda x. (M\sharing{\widetilde{x}}{x})}_u$ first confirms its behavior  along $u$, followed by the reception of a channel $x$.
The channel $x$ provides a linear channel $\linvar{x}$ and an unrestricted channel $\unvar{x}$ for dedicated substitutions of the linear and unrestricted bag components.
This separation is also present in the translation of $ \piencodf{ M\esubst{B}{x}}_u $, for the same reason.

Process $\piencodf{M\, (C \bagsep U)_u}$ consists of synchronizations between the translation of $\piencod{M}_v$ and
$\piencodf{C\bagsep U}_x$:  the translation
of  $C \bagsep U$ evolves when $M$ is an abstraction, say
${\lambda x . (M'\sharing{\widetilde{x}}{x})}$.
The channel $ \linvar{x}$ provides the linear behavior of the bag $C$ while $\unvar{x}$ provides the behavior of $U$. This is done by guarding the translation of $U$ with a server connection: every time a channel synchronizes with it a fresh copy of $U$ is spawned.

As in~\cite{DBLP:conf/aplas/HeuvelPNP23}, non-deterministic choices occur in the translations of   $M\linexsub{C/\widetilde{x}}$ (explicit substitutions) and  $M\sharing{\widetilde{x}}{x}$ (non-empty sharing).
    Roughly speaking, the position  of $\nd$  in the translation of $M\linexsub{C/\widetilde{x}}$  represents the most desirable way of mimicking the fetching of terms from a bag.
This use of $\nd$ is a central idea in our translation: as we explain below,  it allows for appropriate commitment in  \nondt choices, but also for \emph{delayed} commitment when necessary.

For simplicity, we consider explicit substitutions \(M\mkern-3mu\linexsub{C/\widetilde{x}}\) where $C=\bag{\mkern-2mu N_1{,}N_2}$ and \(\widetilde{x}=x_1,x_2\).
The translation \(\piencodf{M \linexsub{C/\widetilde{x}}}_{u}\) uses the processes \(\piencodf{N_i}_{z_i}\), where each $z_i$ is fresh.
First, each bag item confirms its behavior.
Then, a variable~\(x_{i} \in \widetilde{x}\) is chosen non-deterministically;
we ensure that these choices consider all variables.
Note that writing \(\bignd_{x_{i} \in \{x_1,x_2 \}}\bignd_{x_{j}\in \{x_1,x_2\}\setminus x_{i} }\) is equivalent to non-de\-ter\-mi\-nis\-tic\-ally assigning \(x_{i},x_{j} \) to each permutation of \(x_1,x_2\).
The resulting choice involves \(\piencodf{M}_{u}\) with $x_{i}, x_j$ substituted by $z_1, z_2$.
Commitment here is triggered only via synchronizations along $z_1$ or $z_2$; synchronizing with
$\gsome{z_i}{\lfv{N_{i}}};\piencodf{N_{i} }_{ {z_i}}$ then represents fetching   $N_{i}$ from the bag.

The process
\(\piencodf{M\sharing{\widetilde{x}}{x}}_{u}\)
first confirms its behavior along \(x\).
Then it sends a name \(y_i\) on \(x\), on which a failed reduction may be handled.
Next,
the translation confirms again its behavior along \(x\) and non-deterministically receives a reference to an $x_i \in \widetilde{x}$.
Each branch consists of $\piencodf{M\sharing{(\widetilde{x} {\setminus} x_i)}{x}}_u$.
The possible choices are permuted, represented by~\(\bignd_{x_i\in \widetilde{x}}\).
{Synchronizations with $\piencodf{M\sharing{(\widetilde{x} {\setminus} x_i)}{x}}_u$ and bags delay commitment in this choice (we return to this point below).}
The process \(\piencodf{M\sharing{}{x}}_{u}\) is similar but simpler: here the  name \(x\) fails, as it cannot take further elements to substitute.

Process $\piencodf{ M \unexsub{U / {x}}}_u $ consists of  the composition of the translation of $M$ and a server guarding the translation of $U$: in order for $\piencodf{M}_u$ to gain access to $\piencodf{U}_{x_i}$ it must first synchronize with the server channel $\unvar{x}$ to spawn a fresh copy of the translation of $U$.
In case of a failure (i.e., a mismatch between the size of the bag~\(C\) and the number of variables in $M$), our translation ensures that the confirmations of \(C\) will not succeed.
This is how failure in \lamcoldetsh is correctly translated to failure in~\fullclpi.

\subsection{Static Correctness: Type Preservation}
 Our translation preserves types: intersection types in \lamcoldetsh are translated to session types in \fullclpi (\Cref{fig:enc_typesunres}). The translation of types describes how non-deterministic fetches are codified as non-deterministic session protocols. As discussed in \Cref{ss:piTypeSys}, session types effectively abstract away from the behavior of processes, as all branches of a non-deterministic choice use the same typing context.
Thus, it is expected that the translation of types remains unchanged w.r.t. those in~\cite{DBLP:conf/fscd/PaulusN021,DBLP:conf/aplas/HeuvelPNP23}. 
Our translation enjoys \emph{static correctness}: well-formed terms in \lamcoldetsh translate to well-typed processes in \fullclpi. We need the following definition.

\begin{definition}\label{def:enc_sestypfailunres}
    Let $
        \Gamma =  {{x}_1: \sigma_1} , \ldots, {{x}_m : \sigma_m} , {{v}_1: \pi_1} , \ldots ,  {v}_n: \pi_n
    $ be a linear context. Also, consider the unrestricted context
    $\Theta =\unvar{x}[1] : \eta_1 , \ldots , \unvar{x}[k] : \eta_k$.
    Translation  $\piencodf{\cdot}_{\_}$  in \Cref{fig:enc_typesunres} extends to $\Gamma, \Theta$  as follows:
    \begin{align*}
        \piencodf{\Gamma} &= {x}_1 : \with \overline{\piencodf{\sigma_1}} , \ldots ,   {x}_m : \with \overline{\piencodf{\sigma_m}} ,
        {v}_1:  \overline{\piencodf{\pi_1}_{(\sigma, i_1)}}, \ldots ,  {v}_n: \overline{\piencodf{\pi_n}_{(\sigma, i_n)}}\\
        \piencodf{\Theta}&=\unvar{x}[1] : \dual{\piencodf{\eta_1}} , \ldots , \unvar{x}[k] : \dual{\piencodf{\eta_k}}
    \end{align*}
\end{definition}

\begin{figure}[t]
    \begin{mdframed} \mysmall
        \begin{align*}
            \piencodf{\unit} &= \with \onef
            &
           \qquad \piencodf{ \eta } &= {!} \with_{\eta_i \in \eta} \{ i : \piencodf{\eta_i} \}
            \\
            \piencodf{(\sigma^{k} , \eta )   \rightarrow \tau} &= \with( \dual{\piencodf{ (\sigma^{k} , \eta  )  }_{(\sigma, i)}} \ampy \piencodf{\tau})
            &
           \qquad  \piencodf{ (\sigma^{k} , \eta  )  }_{(\sigma, i)} &= \oplus( (\piencodf{\sigma^{k} }_{(\sigma, i)}) \otimes (( \piencodf{\eta}) \otimes (\onef))  )
            \\[1mm]
            \piencodf{ \sigma \wedge \pi }_{(\sigma, i)} & = \oplus(( \with \onef) \ampy ( \oplus  \with (( \oplus \piencodf{\sigma} ) \otimes (\piencodf{\pi}_{(\sigma, i)}))))
            \span \span
            \\
            \piencodf{\omega}_{(\sigma, i)} & = \begin{cases}
                \oplus ((\with \1) \parr (\oplus \with \1))
                &  \text{if $i = 0$}
                \\
                \oplus ((\with \1) \parr (\oplus\, \with ((\oplus \piencodf{\sigma}) \tensor (\piencodf{\omega}_{(\sigma, i-1)}))))
                & \text{if $i > 0$}
            \end{cases}
            \span \span
        \end{align*}
    \end{mdframed}
    \caption{Translation of intersection types into session types  (cf.\ \defref{def:enc_sestypfailunres}).}
    \label{fig:enc_typesunres}
\end{figure}

\begin{restatable}[Type Preservation]{theorem}{thmEncTypePres}\label{t:preservationencode}
    Let $B$ and ${M}$ be a bag and an term in \lamcoldetsh, respectively.
    \begin{enumerate}
        \item If $\Theta ; \Gamma \wfdash B : (\sigma^{k} , \eta )$
        then
        $\piencodf{B}_u \vdash  \piencodf{\Gamma}, u : \piencodf{(\sigma^{k} , \eta )}_{(\sigma, i)} , \piencodf{\Theta}$.

        \item If $\Theta ; \Gamma \wfdash M : \tau$
        then
        $\piencodf{{M}}_u \vdash  \piencodf{\Gamma}, u :\piencodf{\tau} , \piencodf{\Theta}$.
    \end{enumerate}
\end{restatable}

\subsection{Dynamic Correctness Under the Eager Semantics}\label{a:secloose}

To state \emph{dynamic correctness}, we rely on established  notions that (abstractly) characterize \emph{correct translations}~\cite{DBLP:journals/iandc/Gorla10,DBLP:phd/dnb/Peters12a,DBLP:journals/corr/abs-1908-08633}.
A language \({\cal L}=(L,\to)\)  consists of a set of terms $L$ and a reduction relation $\to$ on $L$.
Each language ${\cal L}$ is assumed to contain a success constructor \(\sucs{}\).
A term \(T \in  L\) has {\em success}, denoted \(\succp{T}{\sucs{}}\), when there is a sequence of reductions (using \(\to\)) from \(T\) to a term satisfying success criteria.

Given ${\cal L}_1=(L_1,\to_1)$ and ${\cal L}_2=(L_2,\to_2)$, we seek translations \(\encod{\cdot }{}: {L}_1 \to {L}_2\) that are correct, i.e., translations that satisfy well-known correctness criteria.
The next definition formulates such  criteria.

\begin{definition}\label{d:encCriteria}
    \textbf{(Correct Translation)}
    Let $\mathcal{L}_1= (\mathcal{M}, \shred_1)$ and  $\mathcal{L}_2=(\mathcal{P}, \shred_2)$ be two languages.
    Let $ \asymp_2 $ be an equivalence 
  over $\mcl{L}_2$.
    We use $M,M' $ (resp.\ $P,P' $) to range over terms in $\mcl{M}$  (resp.\ $\mcl{P}$).
  Given a translation $\encod{\cdot }{}: {\cal M}\to {\cal P}$, we define:
    \begin{description}
        \item \textbf{Completeness:}
            For every ${M}, {M}' $ such that ${M} \shred_1^\ast {M}'$, there exists $ P $ such that $ \encod{{M}}{} \shred_2^\ast P \asymp_2 \encod{{M}'}{}$.

        %
        \item \textbf{Weak Soundness:}
            For every $M$ and $P$ such that $\encod{M}{} \shred_2^\ast P$, there exist  $M'$, $P' $ such that $M \shred_1^\ast M'$ and $P \shred_2^\ast P' \asymp_2 \encod{M'}{}  $.

        %
        \item \textbf{Success Sensitivity:}
            For every ${M}$, we have $\succp{M}{\sucs{}}$ if and only if $\succp{\encod{M}{}}{\sucs{}}$.
    \end{description}
    %
\end{definition}


Hence, to prove that our translation of  \lamcoldetsh into \fullclpi
is correct, we need to instantiate Definition~\ref{d:encCriteria} appropriately. 
It turns out that to prove that our translation is correct, we 
need to instantiate  $ \asymp_2 $ with a pre-congruence  on processes, denoted $\premat$,  defined as follows:
\begin{mathpar}
    \inferrule{ }{
        P \premat P
    }
    \and
    \inferrule{
        P_i \premat P'_i
        \\
        {\scriptstyle i \in \{1,2\}}
    }{
        P_1 \nd P_2 \premat P'_i
    }
    \and
    \inferrule{
        P \premat P'
        \\
        Q \premat Q'
    }{
        P   \| Q \premat P'   \| Q'
    }
    \and
    \inferrule{
        P \premat P'
    }{
        \res{ x }   P \premat  \res{ x }  P'
    }
\end{mathpar}
Intuitively, $P \premat Q$ says that $P$ has at least as many branches as $Q$.
We have the following properties, which are instances of those stated in Definition~\ref{d:encCriteria}:

\begin{restatable}[Loose Completeness]{theorem}{thmEncLWCompl}\label{thm:opcompletenessweak}
    If $ {N}\red {M}$ for a well-formed closed $\lamcoldetsh$-term $N$, then there exists $Q$ such that $\piencodf{{N}}_u \redone^* Q$ and $\piencodf{{M}}_u \premat Q $.
\end{restatable}


\begin{restatable}[Loose Weak Soundness]{theorem}{thmEncLWSound}\label{thm:opsoundweak}
    If  $\piencodf{{N}}_u \redone^* Q$ for a well-formed closed $\lamcoldetsh$-term $N$, then there exist ${N}'$ and $Q'$ such that
    (i)~${N} \redone^* {N}'$
    and
    (ii)~$Q \redone^* Q' $ with
    $ \piencodf{{N}'}_u \premat Q'$.
\end{restatable}

\begin{restatable}[Success Sensitivity]{theorem}{thmEncEagerSucc}\label{proof:successsenscetwounres}
    $\succp{{M}}{\sucs{\lambda}}$ iff ${\piencodf{{M}}_u}\succone{\sucs{\pi}}$ for well-formed closed terms $M$.
\end{restatable}

Translation correctness up to $\premat$ thus means that $\redone$ is ``{too eager}'', as it {prematurely commits} to branches.
In contrast, the translation correctness properties established in~\cite{DBLP:conf/aplas/HeuvelPNP23} under the lazy semantics hold by instantiating
$ \asymp_2 $ simply with $\equiv$. 
We thus conclude that moving from the (complex) lazy semantics of~\cite{DBLP:conf/aplas/HeuvelPNP23} to the (simpler) eager semantics of \Cref{ss:pisemantics} has a concrete effect in encodability properties: while translation correctness in the lazy regime is \emph{tight}, it becomes 
\emph{loose} in the eager regime. 

\begin{figure*}[!t]
    \begin{mdframed} \mysmall
        \begin{align*}
            \piencodf{M}_u
            =
            &~
            \res{z_1}(
            \gsome{z_1}{\emptyset}; \piencodf{ \fail^{\emptyset} }_{z_1}
            \| \res{z_2}(
            \gsome{z_2}{y}; \piencodf{ y }_{z_2}
            \| \res{z_3}(
            \gsome{z_3}{\emptyset}; \piencodf{ I }_{z_3}
            \\
            &~ \quad {}
            \| \bignd_{(x_i,x_j,x_k) \in \pi(\{z_1,z_2,z_3\})}
            \res{v}(
            \psome{x_i}; \pfwd{x_i}{v} \| \gsome{v}{v,x_j,x_k}; \pname{v}{z}; (
            \piencodf{ \bag{ x_j \bag{ x_k \ \oneb } } }_z
            \| \pfwd{v}{u}
            ) ) ) ) )
        \end{align*}

        \newlength{\sibldist}\setlength{\sibldist}{1.3cm}
            \mbox{}\hfill
            \begin{tikzpicture}
            [
                level 1/.style = {sibling distance = \sibldist},
                level 2/.style = {sibling distance = 0.5cm},
                level distance = 2.2cm,
                squigly/.style = {decorate,decoration={snake,amplitude=1pt,segment length=6pt,pre length=0pt,post length=2pt}}
            ]

                \node {$\piencodf{M}_u$}
                    child { node {$\piencodf{N_1}_u$}
                        edge from parent [->, squigly, cblGreen] node [above, xshift=-.5ex, pos=1.0, text=black] {\scriptsize $\ast$}}
                    child { node {$\piencodf{N_2}_u$}
                        edge from parent [->, squigly, cblGreen] node [left, yshift=1.0ex, xshift=.1ex, pos=1.0, text=black] {\scriptsize $\ast$}}
                    child { node {$\piencodf{N_3}_u$}
                        edge from parent [->, squigly, cblGreen] node [above, xshift=.5ex, pos=1.0, text=black] {\scriptsize $\ast$}};
            \end{tikzpicture}
            \hfill
            \textcolor{gray}{\vrule}
            \hfill
            \begin{minipage}[t]{11.6cm}
                \centering
                \begin{tikzpicture}
                [
                    level 1/.style = {sibling distance = 2.0cm},
                    level 2/.style = {sibling distance = 0.5cm},
                    level distance = 1.1cm
                ]

                    \node {$\piencodf{ M }_u$ }
                        child { node {$P_1(z_2,z_3)$}
                            child { node [text=black] {$\piencodf{N_1}_u$}
                                edge from parent [draw=none] node [draw=none] {$\premattwo$}}
                            edge from parent [->, cblRed] node [above, pos=1.0, text=black] {\scriptsize $\ast$}}
                        child {node {$P_1(z_3,z_2)$}
                            child { node [text=black] {$\piencodf{N_1}_u$}
                                edge from parent [draw=none] node [draw=none] {$\premattwo$}}
                            edge from parent [->, cblRed] node [above, pos=1.0, text=black] {\scriptsize $\ast$}}
                        child {node {$P_2(z_1,z_3)$}
                            child { node [text=black] {$\piencodf{N_2}_u$}
                                edge from parent [draw=none] node [draw=none] {$\premattwo$}}
                            edge from parent [->, cblRed] node [above, xshift=-.5ex, pos=1.0, text=black] {\scriptsize $\ast$}}
                        child {node {$P_2(z_3,z_1)$}
                            child { node [text=black] {$\piencodf{N_2}_u$}
                                edge from parent [draw=none] node [draw=none] {$\premattwo$}}
                            edge from parent [->, cblRed] node [above, xshift=.5ex, pos=1.0, text=black] {\scriptsize $\ast$}}
                        child {node {$P_3(z_1,z_2)$}
                            child { node [text=black] {$\piencodf{N_3}_u$}
                                edge from parent [draw=none] node [draw=none] {$\premattwo$}}
                            edge from parent [->, cblRed] node [above, pos=1.0, text=black] {\scriptsize $\ast$}}
                        child {node {$P_3(z_2,z_1)$}
                            child { node [text=black] {$\piencodf{N_3}_u$}
                                edge from parent [draw=none] node [draw=none] {$\premattwo$}}
                            edge from parent [->, cblRed] node [above, pos=1.0, text=black] {\scriptsize $\ast$}};
                \end{tikzpicture}

                The processes above are as follows:
                \begin{align*}
                    Q(a,b)
                    &=
                    \gsome{v}{u,a,b}; \pname{v}{z}; ( \piencodf{ \bag{a \ \bag{ b \ \oneb}} }_z \| \pfwd{v}{u} )
                    \\[-1mm]
                    P_1(a,b)
                    &=
                    \res{z_2}( \gsome{z_2}{y}; \piencodf{ y }_{z_2} \| \res{z_3}( \gsome{x_3}{\emptyset}; \piencodf{ I }_{z_3} \| \res{v}( \piencodf{ \fail^{\emptyset} }_{v} \| Q(a,b) ) ) )
                    \\[-1mm]
                    P_2(a,b)
                    &=
                    \res{z_1}( \gsome{z_1}{\emptyset}; \piencodf{ \fail^{\emptyset} }_{z_1} \| \res{z_3}( \gsome{x_3}{\emptyset}; \piencodf{ I }_{z_3} \| \res{v}( \piencodf{ y }_{v} \| Q(a,b) ) ) )
                    \\[-1mm]
                    P_3(a,b)
                    &=
                    \res{z_1}( \gsome{z_1}{\emptyset}; \piencodf{ \fail^{\emptyset} }_{z_1} \| \res{z_2}( \gsome{z_2}{y}; \piencodf{ y }_{z_2} \| \res{v}( \piencodf{ I }_{v} \| Q(a,b) ) ) )
                \end{align*}
            \end{minipage}
            \hfill\mbox{}
    \end{mdframed}
    \caption{Example~\ref{ex:looseTight}: Reductions of  $\piencodf{M}_u$ under $\redtwo_S$ and  $\redone$. We write `$\pi(X)$' for the permutations of set $X$.}
    \label{fig:eager_m_red}\label{f:eager_v_lazy}
\end{figure*}

\begin{example}
\label{ex:looseTight}
    To further contrast commitment in eager and lazy semantics (and their effect on the translation's correctness), recall from Example~\ref{ex:lambdaRed} the term $M$~\eqref{eq:lin_cons_sub2} and the three branching reductions from $M$ to $N_1$, $N_2$ and $N_3$.
    \Cref{f:eager_v_lazy}   depicts a side-by-side comparison of the reductions of   $\piencodf{M}_u$ under the lazy ($\redtwo$) and eager ($\redone$) semantics. We omit the translation of the empty unrestricted bag $\piencodf{\unexsub{\oneb^!/x}}_u$ in the translations of  $\piencodf{M}_u$ and each of the  $\piencodf{N_i}_u$, since it does not add any insight.
    In the figure, $\redtwo^\ast$ and $\redone^\ast$ denote the reflexive, transitive closures of $\redtwo$ and $\redone$, respectively.

    Under \redtwo there are three  reduction paths, each resulting directly in the translation of one of $N_1,N_2,N_3$: after the first choice, the following choices are preserved.
    In contrast, under \redone  there are six  reduction paths, each resulting in a process that relates to the translation of one of $N_1,N_2,N_3$ through $\premat$: after the first choice for an item from the bag is made, the semantics commits to choices for the other items.
\end{example}

\section{Comparing Lazy and Eager Semantics via Behavioral Equivalences}
\label{s:compare}
We now compare \redtwo and \redone  \emph{independently from \lamcoldetsh} by resorting to \emph{behavioral equivalences}.
We define a simple behavioral notion of equivalence on \fullclpi processes, parametric in \redtwo or \redone; then, we prove that there are classes of processes that are equal with respect to \redtwo, but incomparable with respect to \redone (Theorem~\ref{t:piBisim}).
A key ingredient is the following notion of observable on processes:

\begin{definition}\label{d:readyPrefix}
    A process $P$ has a \emph{ready-prefix} $\alpha$, denoted $P \readyPrefix{\alpha}$, iff there exist $\pctx{N},P'$ such that $P \equiv \pctx{N}[\alpha; P']$.
\end{definition}


\begin{definition} \label{d:readyPrefixBisim}
    \textbf{(Ready-Prefix Bisimilarity)}
    A relation $\mathbb{B}$ on \fullclpi processes is a \emph{(strong) ready-prefix bisimulation with respect to \redtwo} if and only if, for every $(P,Q) \in \mathbb{B}$,
    \begin{enumerate}
        \item
            For every $P'$ such that $P \redtwo P'$, there exists $Q'$ such that $Q \redtwo Q'$ and $(P',Q') \in \mathbb{B}$;

        \item
            For every $Q'$ such that $Q \redtwo Q'$, there exists $P'$ such that $P \redtwo P'$ and $(P',Q') \in \mathbb{B}$;

        \item
            For every $\alpha \relalpha \beta$, $P \readyPrefix{\alpha}$ if and only if $Q \readyPrefix{\beta}$.
    \end{enumerate}
    $P$ and $Q$ are \emph{ready-prefix bisimilar with respect to \redtwo}, denoted $P \readyPrefixBisim{L} Q$, if there exists a relation $\mathbb{B}$ that is a
    ready-prefix bisimulation with respect to \redtwo such that $(P,Q) \in \mathbb{B}$.

    A \emph{(strong) ready-prefix bisimulation with respect to \redone} is defined by replacing every occurrence of `$\redtwo$' by `$\redone$' in the definition above.
    We write $P \readyPrefixBisim{E} Q$ if $P$ and $Q$ are \emph{ready-prefix bisimilar with respect to~\redone}.
\end{definition}

Ready-prefix bisimulation can highlight a significant difference between the behavior induced the lazy and eager semantics.
To illustrate this, we consider session-typed implementations of a vending machine.
\newcommand{\myMachine}{VM}
\newcommand{\myInterface}{IF}

\begin{example}\label{ex:prefixReadyBisim}
    \textbf{(Two Vending Machines)}
    Consider vending machines $\sff{\myMachine}_1$ and $\sff{\myMachine}_2$ consisting of three parts:
    (1)~an interface, which interacts with the user to send money and choose between coffee~($\sff{c}$) and tea~($\sff{t}$);
    (2)~a brewer, which produces either beverage;
    (3)~a system, which collects the money and forwards the user's choice to the brewer.
    An \fullclpi specification follows (below $\textup{€}$ and $\textup{€}2$ stand for names):
    \begin{align*}
        \sff{\myMachine}_1 &:= \res{x} \big( \sff{\myInterface}_1 \| \res{y} ( \sff{Brewer} \| \sff{System} ) \big) \qquad 
        \sff{\myMachine}_2 := \res{x} \big( \sff{\myInterface}_2 \| \res{y} ( \sff{Brewer} \| \sff{System} ) \big)
            \\ \displaybreak[1]
        \sff{\myInterface}_1 &:= \pname{x}{\textup{€}2}; \big( \pclose{\textup{€}2} \| ( \psel{x}{\sff{c}}; \pclose{x} \nd \psel{x}{\sff{t}}; \pclose{x} ) \big) \qquad 
        \sff{\myInterface}_2 := \pname{x}{\textup{€}2}; ( \pclose{\textup{€}2} \| \psel{x}{\sff{c}}; \pclose{x} ) \nd \pname{x}{\textup{€}2}; ( \pclose{\textup{€}2} \| \psel{x}{\sff{t}}; \pclose{x} )
                    \\ \displaybreak[1]
        \sff{System} &:= \gname{x}{\!\textup{€}}; \gsel{x} \!\left\{\!
            \begin{array}{@{}l@{}}
                \sff{c} : \psel{y}{\sff{c}}; \gclose{x}; \gclose{\textup{€}}; \pclose{y},
                \\
                \sff{t} : \psel{y}{\sff{t}}; \gclose{x}; \gclose{\textup{€}}; \pclose{y}
            \end{array}
        \!\right\} \qquad 
        \sff{Brewer} := \gsel{y} \{ \sff{c} : \gclose{y}; \sff{Brew}_{\sff{c}}, \sff{t} : \gclose{y}; \sff{Brew}_{\sff{t}} \}
    \end{align*}
    where $\sff{Brew}_{\sff{c}} \vdash \emptyset$, $\sff{Brew}_{\sff{t}} \vdash \emptyset$, such that $\sff{\myMachine}_1 \vdash \emptyset$, $\sff{\myMachine}_2 \vdash \emptyset$.

The  machines $\sff{\myMachine}_1$ and $\sff{\myMachine}_2$ are based on two different implementations of the interface. Machine $\sff{\myMachine}_1$ uses 
    $\sff{\myInterface}_1$, which sends the money and then chooses coffee or tea.
    Machine $\sff{\myMachine}_2$ uses
    $\sff{\myInterface}_2$, which chooses sending the money and then requesting coffee, or sending the money and then requesting tea.

    We have $\sff{\myMachine}_1 \nreadyPrefixBisim{E} \sff{\myMachine}_2$: the eager semantics distinguishes  the machines; e.g., $\sff{\myInterface}_1$ has a single money slot, a button for coffee, and another button for tea, whereas $\sff{\myInterface}_2$ has two money slots, one for coffee, and another for tea.
    In contrast, the machines are indistinguishable under the lazy semantics:
    $\sff{\myMachine}_1 \readyPrefixBisim{L} \sff{\myMachine}_2$.
\end{example}

Example~\ref{ex:prefixReadyBisim} highlights a difference in behavior between \redtwo and \redone when a moment of choice is subtly altered.
The following theorem captures this distinction; we need an auxiliary definition.
    Let $\relalpha$ denote the least relation on prefixes   defined by: 
    (i)~$\pname{x}{y} \relalpha \pname{x}{z}$, (ii)~$\gname{x}{y} \relalpha \gname{x}{z}$, and (iii)~$\alpha \relalpha \alpha$ otherwise.

\begin{restatable}{theorem}{thmPiBisim}\label{t:piBisim}
    Take $R \equiv \pctx{N}[\alpha_1; (P \nd Q)] \vdash \emptyset$ and $S \equiv \pctx{N}[\alpha_2; P \nd \alpha_3; Q] \vdash \emptyset$, where $\alpha_1 \relalpha \alpha_2 \relalpha \alpha_3$ and $\alpha_1,\alpha_2,\alpha_3$ require a continuation.
    Suppose that $P \nreadyPrefixBisim{L} Q$ and $P \nreadyPrefixBisim{E} Q$.
    Then (i)~$R \readyPrefixBisim{L} S$ but (ii)~$R \nreadyPrefixBisim{E} S$.
\end{restatable}

%
%


\section{Conclusion}
We studied how to reconcile non-deterministic choice and linearity in a  session-typed $\pi$-calculus. 
We extended and complemented the results in~\cite{DBLP:conf/aplas/HeuvelPNP23}, which were based on a lazy semantics for expressing commitment in non-deterministic choices.
Our central contribution is an eager semantics that more directly expresses such commitment while respecting linearity.
We confirmed that this eager semantics fits well with the session type discipline: both type preservation and deadlock-freedom properties are still ensured in the eager regime. 
We also showed the expressivity of our typed model by giving a correct translation of a resource  $\lambda$-calculus, which extends the analogous results in~\cite{DBLP:conf/aplas/HeuvelPNP23} with  unrestricted resources.
We compared the two semantics in two ways: (i)~via the  correctness properties they induce for the translation of resource $\lambda$-calculi, and (ii)~via a simple behavioral equivalence that captures different moments of choice.
Based on our results, we conclude that eager and lazy semantics are both worth studying, as each of them has merits and shortcomings: the lazy semantics is complex but admits fine-grained observations; the new eager semantics has a simpler definition but induces commitment that can be sometimes too premature.

\paragraph{Acknowledgments}
 We are grateful to the anonymous reviewers for useful comments on previous versions of this paper. 
 
 This research has been supported by the Dutch Research Council (NWO) under project No. 016.Vidi.189.046 (‘Unifying Correctness for Communicating Software’) and by the EPSRC ‘VeTSpec: Verified Trustworthy Software Specification’ (EP/R034567/1), ‘Session Types for Reliable Distributed Systems (STARDUST)’ (EP/T014709/2), and ‘Verified Simulation for Large Quantum Systems (VSL-Q)’ (EP/Y005244/1).


\bibliographystyle{./entics}
\bibliography{references}
%

\end{document}